\documentclass[prd,amsmath,superscriptaddress,twocolumn]{revtex4}

\pdfoutput=1

\usepackage{graphicx} 
\usepackage{bmpsize}
\usepackage{amsmath}
\usepackage{amssymb}
\usepackage{graphics}
\usepackage{hyperref}
\usepackage{mathrsfs}
\usepackage{bm}
\usepackage[usenames,dvipsnames]{color}

\newcommand{\Pdd}{P_{\delta\delta}}
\newcommand{\Pdt}{P_{\delta\theta}}

\newcommand{\Ptt}{P_{\theta\theta}}

\newcommand{\deltag}{\delta_\mathrm{g}}
\newcommand{\thetg}{\theta_\mathrm{g}}
\newcommand{\Pgdd}{P_{\deltag \deltag}}
\newcommand{\Pgdt}{P_{\deltag \thetg}}
\newcommand{\Pgtt}{P_{\thetg \thetg}}
\newcommand{\bd}{b_\delta}
\newcommand{\bdd}{b_{\delta^2}}
\newcommand{\bss}{b_{s^2}}
\newcommand{\btnl}{b_{3\mathrm{nl}}}
\newcommand{\bv}{b_v}
\newcommand{\bnu}{b_\mathrm{n}}
\newcommand{\Fspt}{F_\mathrm{S}^{(2)}}
\newcommand{\Gspt}{G_\mathrm{S}^{(2)}}

\newcommand{\Pbis}{P^\mathrm{B}}
\newcommand{\Pbisj}[1]{P^\mathrm{B}_{#1}}

\newcommand{\Ptri}{P^\mathrm{T}}
\newcommand{\Ptrij}[1]{P^\mathrm{T}_{#1}}
\newcommand{\Qlabc}{Q_{abc}^{(\ell)}}
\newcommand{\QQ}[2]{Q^{(#1)}_{#2}}
\newcommand{\Rlabc}{R_{abc}^{(\ell)}}

\newcommand{\Camb}{{\tt{CAMB}}}

\newcommand{\CosmoMC}{{\tt{CosmoMC}}}
\newcommand{\deltacb}{\delta_\mathrm{CB}}

\newcommand{\deltanu}{\delta_\nu}
\newcommand{\Dirac}{\delta_\mathrm{D}}
\newcommand{\fcb}{f_\mathrm{CB}}
\newcommand{\Ffog}{F_\mathrm{fog}}
\newcommand{\fnu}{f_\nu}
\newcommand{\Hc}{{\mathcal H}}
\newcommand{\dk}{\delta^{(\mathrm{K})}}

\newcommand{\ns}{n_\mathrm{s}}
\newcommand{\omegab}{\omega_\mathrm{b}}

\newcommand{\Omegabo}{\Omega_{{\mathrm b}0}}
\newcommand{\omegac}{\omega_\mathrm{c}}

\newcommand{\Omegaco}{\Omega_{{\mathrm c}0}}
\newcommand{\Omegam}{\Omega_\mathrm{m}}
\newcommand{\Omegamo}{\Omega_{\mathrm{m}0}}
\newcommand{\omegam}{\omega_\mathrm{m}}
\newcommand{\omeganu}{\omega_\nu}

\newcommand{\Omeganuo}{\Omega_{\nu 0}}

\newcommand{\Pleg}{{\mathscr P}}

\newcommand{\plik}{{\tt{plik}}}
\newcommand{\Plin}{P_\mathrm{lin}}

\newcommand{\Pnu}{P_\nu}
\newcommand{\Peffnu}{P_\mathrm{eff}^{(\nu)}}
\newcommand{\Peffnun}{P_\mathrm{eff}^{(\nu,n)}}
\newcommand{\redtime}{{\tt{redTime}}}

\newcommand{\xdd}{x^{(\mathrm{dd})}}
\newcommand{\xdc}{x^{(\mathrm{dc})}}
\newcommand{\xdnu}{x^{(\mathrm{d}\nu)}}
\newcommand{\xdN}{x^{(\mathrm{dN})}}
\newcommand{\xcc}{x^{(\mathrm{cc})}}
\newcommand{\xcnu}{x^{(\mathrm{c}\nu)}}
\newcommand{\xcN}{x^{(\mathrm{cN})}}
\newcommand{\xnunu}{x^{(\nu\nu)}}
\newcommand{\xnuN}{x^{(\nu\mathrm{N})}}
\newcommand{\xNN}{x^{(\mathrm{NN})}}

\newcommand{\kfs}{k_\mathrm{fs}}
\newcommand{\knr}{k_\mathrm{nr}}
\newcommand{\Neff}{{\mathcal N}_\mathrm{eff}}
\newcommand{\vth}{v_\mathrm{th}}

\begin{document}

\title{Neutrino mass and dark energy constraints from redshift-space distortions}
\author{Amol Upadhye}
\affiliation{Department of Physics, 
  University of Wisconsin--Madison, 
  1150 University Avenue, Madison, WI 53706}
\affiliation{School of Physics, 
  The University of New South Wales, 
  Sydney NSW 2052, Australia}
\date{\today}

\begin{abstract}
  Cosmology in the near future promises a measurement of the sum of neutrino masses $\sum m_\nu$, a fundamental Standard Model parameter, as well as substantially-improved constraints on the dark energy. We use the shape of the BOSS redshift-space galaxy power spectrum, in combination with CMB and supernova data, to constrain the neutrino masses and the dark energy.   Essential to this calculation are several recent advances in non-linear cosmological perturbation theory, including fast Fourier transform methods, redshift space distortions, and scale-dependent growth.  Our $95\%$ confidence upper bound $\sum m_\nu < 180$~meV degrades substantially to $\sum m_\nu < 540$~meV when the dark energy equation of state and its first derivative are also allowed to vary, representing a significant challenge to current constraints.  We also study the impact of additional galaxy bias parameters, finding that a greater allowed range of scale-dependent  bias only slightly shifts the preferred $\sum m_\nu$, weakens its upper bound by $\approx 20\%$, and has a negligible effect on the other cosmological parameters.
\end{abstract}

\maketitle

\section{Introduction}
\label{sec:introduction}

Cosmology over the last twenty years has established itself as an important probe of fundamental physics. A standard cosmological picture has emerged in which the seeds of the largest-scale gravitationally-bound structures are approximately Gaussian, adiabatic density perturbations with a slightly red-tilted spectrum in a universe that is nearly spatially flat.  Baryonic matter makes up approximately $5\%$ of the total energy density.  Another $\approx 25\%$ is ``dark matter,'' which does not interact with photons.  This is mostly ``cold,'' or non-relativistic, though $\sim 1\%$ of it consists of a marginally relativistic ``warm'' massive neutrino fluid.  ``Dark energy,'' a mysterious negative-pressure fluid, makes up the remainder of the energy density.  Confirming this basic picture are data from the cosmic microwave background (CMB)~\cite{Adam_2016a,Ade_2016m,Baxter_2014,Benson_2014,Sherwin_2016,Louis_2016}, Type IA supernovae~\cite{Betoule_2014}, galaxy redshift surveys~\cite{Beutler_2014,Beutler_2014b,Beutler_2017,Alam_2016,Hinton_2017}, weak gravitational lensing~\cite{Kwan_2017,Aihara_2017,Joudaki_2017,deJong_2017}, the Hubble diagram of cosmic distance measurements~\cite{Riess_2016}, and the ``forest'' of Lyman-$\alpha$ lines in quasar spectra~\cite{Bautista_2017}.

Nevertheless, as the data have improved over the past several years, a few $2\sigma-3\sigma$ inconsistencies have emerged among the data sets. The Hubble parameter $H_0 \approx 73$~km/sec/Mpc from local distance measurements~\cite{Riess_2016} is about $3\sigma$ higher than the value $H_0 \approx 67$~km/sec/Mpc measured cosmologically~\cite{Ade_2016m}.  Gravitational lensing of the CMB appears $15\%$ higher than predicted by General Relativity, a $> 2\sigma$ discrepancy~\cite{Aghanim_2016b}.  Galaxy-galaxy lensing prefers significantly lower values of either the cold matter density or the amplitude of density perturbations~\cite{Leauthaud_2017}. These tensions could indicate systematic biases which would have to be understood in order for  progress to be made~\cite{Keenan_2014,Addison_2016}, but could also be early indicators of new physics\cite{Wyman_2014,Bernal_2016,DiValentino_2017}.

In this article we analyze  the Planck CMB power spectrum of Ref.~\cite{Adam_2016a} as well as the Fourier space BOSS power spectra of Refs.~\cite{Beutler_2014,Beutler_2014b,Beutler_2017}, two data sets which appear not to have any significant tensions, and for our dark energy analyses we also include the Joint Likelihood Analysis of supernova data presented in Ref.~\cite{Betoule_2014}.  This article has three main aims.  First, we constrain the sum of neutrino masses, $\sum m_\nu$, a fundamental Standard Model parameter whose first significant detection will likely come from cosmology~\cite{Font-Ribera_2014,Allison_2015}.  Though we use the BOSS Data Release 11 redshift-space galaxy power spectrum of Ref.~\cite{Beutler_2014,Beutler_2014b}, our best-fit $\sum m_\nu$ is somewhat lower than found in those references, a shift which may be attributed to our different handling of the scale-dependent suppression of density fluctuations in massive neutrino models, as well as to our use of more recent CMB data.

Second, we constrain the time-dependent dark energy equation of state $w(z) = P_\mathrm{de}/\rho_\mathrm{de}$ in models with and without variable $\sum m_\nu$.  Comparing these two analyses allows us to assess the impact of $\sum m_\nu$ on $w(z)$ constraints, and vice versa.  In particular, allowing $w$ and its late-time derivative to vary worsens the upper bound on $\sum m_\nu$ by a factor of $\approx 3$.

Third, we thoroughly investigate the dependence of the neutrino mass constraint on the modeling of scale-dependent galaxy bias and choice of data sets.  Comparing the perturbative bias parameterization of McDonald and Roy, Ref.~\cite{McDonald_Roy_2009}, to the galaxy power spectra of Ref.~\cite{Kwan_etal_2015} based upon N-body simulations, we find a broad agreement at the $1\%$ level over the range of scales relevant to the BOSS data.  Adding more density bias parameters to the minimal bias model has a negligible effect on the mean $\sum m_\nu$ but worsens its upper bound by $20\%$.  Reducing the maximum wave number considered, switching from BOSS DR11 to DR12 data, and modifying the neutrino contribution to the galaxy power spectrum, also weaken neutrino mass constraints.  The remaining cosmological parameters are quite robust.

This work takes advantage of several recent theoretical developments.  Time-Renormalization Group perturbation theory was designed in Refs.~\cite{Pietroni_2008,Lesgourgues_etal_2009} for cosmological models with scale-dependent linear growth, including massive neutrino models.  It was compared with N-body dark matter simulations in Refs.~\cite{Carlson_White_Padmanabhan_2009,Upadhye_2014,Upadhye_2016}, the last of which extended it to redshift space in the code \redtime. Meanwhile, the Fast Fourier Transform (FFT) techniques of Refs.~\cite{Schmittfull_2016,McEwen_2016,Fang_2017}, named FAST-PT in Ref.~\cite{McEwen_2016}, speed up perturbation theory integrals considerably. Here we have used FAST-PT to speed up \redtime~by a factor of over forty.  Its new running time is a few seconds on an eight-processor desktop machine, comparable to the running time of \Camb.  Since the technical details are not necessary for understanding our results, we defer their discussion to the appendices, along with descriptions of our implementation of the bias model of Ref.~\cite{McDonald_Roy_2009} and the BOSS DR11 likelihood function of Ref.~\cite{Beutler_2014b}.

The article is organized as follows.  Section~\ref{sec:background} provides overviews of massive neutrino cosmology, redshift-space distortions, and galaxy bias.  The three data sets we use are summarized in Sec.~\ref{sec:data_sets} along with our Monte Carlo Markov Chain analysis.  Our results are tabulated and described in Sec.~\ref{sec:results_and_discussion}, and Sec.~\ref{sec:conclusions} concludes.  Three appendices provide more detail on our galaxy bias implementation, the FAST-PT enhancement of the \redtime~redshift-space perturbation code, and our implementation of the galaxy survey likelihood allowing for rapid marginalization over the bias parameters.

\section{Background}
\label{sec:background}

\subsection{Massive neutrinos and structure formation}
\label{subsec:massive_neutrinos_and_structure_formation}

Massive neutrinos behave as a warm component of the dark matter, which clusters like cold matter on the largest scales but whose thermal velocity exceeds the escape velocities of smaller-scale cosmic structures.  References~\cite{Lesgourgues_Pastor_2006,Lesgourgues_Pastor_2012} provide thorough reviews of the cosmological impacts of massive neutrinos, which we summarize here.

During Big Bang Nucleosynthesis (BBN), Standard Model neutrinos are ultra-relativistic, with distribution function $f(\vec p,\mu,T) = 1 / [\exp((p-\mu)/T) + 1].$  In the simplest models, $\mu/T$ is undetectably small, and we neglect it here. When the weak interaction rate $\Gamma_\nu = \left< \sigma_\nu n_\nu \right> \sim G_\mathrm{F}^2 T^5$ coupling neutrinos to other particles drops below the Hubble expansion rate $H$, neutrinos fall out of equilibrium with the rest of the radiation. In practice, this occurs around $T = 1$~MeV. Soon afterwards, $H$ drops below the electron mass, and electron-positron annihilation heats the photon gas to a temperature $\approx (11/4)^{1/3}$ times the neutrino temperature in the approximation of instantaneous neutrino decoupling.  The total radiation energy density after electron-positron annihilation is parameterized
\begin{equation}
  \rho_\mathrm{rad}
  =
  \left[1 + \frac{7}{8} \left(\frac{4}{11}\right)^{4/3} \Neff \right]
  \frac{\pi^2}{15} T_\gamma^4
\end{equation}
where $T_\gamma$ is the time-dependent photon temperature and $\Neff$ the effective number of neutrinos.  Since neutrino decoupling is not exactly instantaneous, neutrinos do absorb some energy from electron-positron annihilation, raising their temperature by $\approx 0.4\%$ above the instantaneous approximation.  This is accommodated by setting the Standard Model value of $\Neff = 3.046$ rather than 3.

Thus far we have discussed neutrinos as effectively massless objects.  In the matter-dominated era, neutrino masses $\lesssim 1$~eV can affect late-time large-scale cosmic structure in three broad ways:
\begin{enumerate}
\item slightly changing the redshift of matter-radiation equality, and the
  resulting turnover in the matter power spectrum;
\item suppressing the total matter power spectrum at small scales by not
  clustering;
\item suppressing the cold matter power at small scales by not
  sourcing CDM+baryon clustering.
\end{enumerate}
We define the neutrino ``free streaming'' scale as
\begin{equation}
  \kfs
  =
  \sqrt{4\pi G \bar\rho} \, a / \vth
  =
  \sqrt{3/2} \, a H / \vth
\end{equation}
with thermal velocity $\vth \approx 1$ for a relativistic neutrino and $\vth = \left< p \right> / m_\nu \approx 3.15 T_\nu/m_\nu$ for a neutrino of mass $m_\nu \gg T_\nu$.  At $z=0$, $\kfs \approx 0.8 (m / 1\mathrm{ eV})~h/$Mpc.  For neutrinos $m_\nu \lesssim 1$~eV which become non-relativistic in the matter-dominated era, the free-streaming wave number reaches a minimum
\begin{equation}
  \knr
  \approx
  0.018 \sqrt{m_\nu \Omegamo / (1~\mathrm{eV})}~h/\mathrm{Mpc}
\end{equation}
below which neutrinos cluster like CDM.

At larger wave numbers, neutrino clustering depends in a more complicated way on time and scale, with the limiting behavior being a negligible neutrino density contrast $\deltanu \ll \deltacb$ at $k \gg \knr$.  (Here, subscripts $\nu$ and $\mathrm{CB}$ refer, respectively, to neutrinos and the combined CDM+baryon fluid.)  For a neutrino fraction
\begin{eqnarray}
  \fnu
  &=&
  \Omeganuo / \Omegamo = \omeganu / \omegam,
  \\
  \omeganu
  &\approx&
  \frac{\sum m_\nu}{93.14~\mathrm{eV}},
\end{eqnarray}
this deficit of neutrino clustering lowers the total matter power spectrum at small scales by a factor $\fcb^2 = (1-\fnu)^2$.  Additionally, since neutrino thermal velocities $\vth / c \approx 5\times 10^{-4} (1~\mathrm{eV}/m_\nu)$ today are near the escape velocities $10^{-4} - 10^{-3}$ of typical cosmic structures, neutrinos tend not to be captured by such structures, leading to the suppression of their gravitational potentials and hence $\deltacb$.  In a linear, matter-dominated universe at small scales $k \gg \knr$, the CDM growth factor is suppressed by a factor $a^{-3 \fnu / 5}$.  At $z=0$, the combination of these effects and non-linear clustering was shown in N-body simulations to reduce the small-scale power by a fraction $\Delta P / P \approx -10 \fnu$, slightly greater in magnitude than the linear-theory reduction~\cite{Lesgourgues_Pastor_2012}.

Henceforth we consider the minimal neutrino parameter space.  We fix $\Neff = 3.046$ and do not consider additional ``sterile'' neutrino species.  Since cosmological data are far from being able to distinguish among the three Standard Model neutrino species, we approximate them as degenerate in mass, and characterized entirely by the parameter $\omeganu$ proportional to the sum of their masses~\cite{Hamann_2012}.

\subsection{Redshift-space distortions}
\label{subsec:redshift-space_distortions}

The observable which we use to characterize the large-scale distribution of galaxies is the redshift-space power spectrum $P_s(\vec k)$, the Fourier transform of the two-point correlation function of the redshift-space density field. Although the density and power spectrum are in principle gauge-dependent objects whose horizon-scale behavior requires a careful consideration of General Relativistic effects~\cite{Yoo_2014,Gong_2017}, current galaxy surveys are insensitive to such effects.  Here we describe non-linear structure formation in the subhorizon regime using Newtonian gravity in a box expanding at the Hubble rate.  Furthermore, we neglect the vorticity of fluid velocity fields, $\vec \nabla \times \vec v = 0$, allowing us to describe matter clustering using the scalar variables $\delta = (\rho - \bar \rho)/\bar\rho$ and $\theta = -\vec \nabla \cdot \vec v / \Hc$, with $\Hc = a H$ and $\bar \rho$ the spatial average of the density $\rho$~\cite{Pueblas_Scoccimarro_2009}.  Finally, we consider only spatially flat universes, $\Omega_K = 0$.

Redshift-space distortions, apparent anisotropies in the measured power that align with the line of sight, are caused by an imperfect mapping between the observed redshift of an object and its inferred line-of-sight distance~\cite{Sargent_Turner_1977}.  In a spatially flat and perfectly homogeneous universe, the comoving distance to an object at redshift $z$ is $\chi(z) = \int_0^z dz' / H(z')$.  In an inhomogeneous universe, an object with a peculiar velocity pointing towards the observer will have a smaller redshift $z$ than a nearby object with no peculiar velocity, and its actual distance will be greater than $\chi(z)$.  Treating $\chi(z)$ as its position results in an apparent anisotropy in the power spectrum which contains information about the peculiar velocity field and the gravitational potential which sources it.

The discussion of linear redshift-space distortions in Ref.~\cite{Kaiser_1987} is instructive.  The Jacobian determinant of the transformation from comoving coordinate $\vec x$ to apparent ``redshift-space'' coordinate $\vec s$ is $J = \left|d^3x / d^3s\right| = 1 / (1 + \partial_x \vec v \cdot \hat r / \Hc)$ with $\hat r$ the line-of-sight direction; note that the second equality above assumes the fluid approximation, which neglects stream crossing.  In terms of the density $\delta(\vec k, t)$ and velocity divergence $\theta(\vec k,t)$, the redshift-space density is $\delta_s = (\delta + \mu^2\theta)J$, where $\mu = \hat k \cdot \hat r$ is the cosine of the line-of-sight angle. In the linear theory of Ref.~\cite{Kaiser_1987}, the resulting power spectrum is $P_{s,\mathrm{lin}}(k,\mu) = (1 + f \mu^2)^2 \Plin(k)$, where $f = d \log(D) / d \log(a)$ and $D$ is the linear growth factor.

Non-linear corrections to $P_{s,\mathrm{lin}}(k,\mu)$ include higher-order terms as well as a streaming factor to account for the ``finger of god'' effect, the apparent redshift-space elongation of virialized objects~\cite{Jackson_1972,Scoccimarro_2004,Taruya_etal_2010,Fisher_1995,Kwan_2012}:
\begin{equation}
  \!\!P_s
  =
  \Ffog(f k \sigma_v \mu)
  \!\left[
    \Pdd \!+\! 2 \mu^2 \Pdt \!+\! \mu^4 \Ptt
    \!+\! \Pbis \!+\! \Ptri
    \right].
\end{equation}
Here $\sigma_v$ is a length scale associated with the velocity dispersion, and $\Ffog$ falls off rapidly for $f k \sigma_v \mu \gg 1$.  We choose a Lorentzian function, $\Ffog(x) = 1 / (1 + x^2)$, tested against N-body simulations in Ref.~\cite{Upadhye_2016}, and leave $\sigma_v$ a free parameter to be fit to the data.  $\Pbis(k,\mu)$ and $\Ptri(k,\mu)$, which respectively depend on the three-point and four-point correlation functions, were introduced in Ref.~\cite{Taruya_etal_2010} and are computed in Appendix~\ref{sec:time-rg_with_fast-pt}.

In this work we compute $P_s(k,\mu)$ using the \mbox{\redtime} redshift-space one-loop Time-Renormalization Group (Time-RG) code of Ref.~\cite{Upadhye_2016}, sped up substantially through the Fast Fourier Transform methods of Refs.~\cite{McEwen_2016,Fang_2017}, with input linear power spectra generated using the \mbox{\Camb} code of Ref.~\cite{Lewis_2000}.  Time-RG perturbation theory uses the irrotational continuity and Euler equations to generate an infinite tower of evolution equations for higher-order correlation functions, which is truncated by neglecting the connected part of the four-point function~\cite{Pietroni_2008}.  Since it integrates the equation of motion separately for each wave number $k$, it automatically accounts for the scale-dependent growth in massive neutrino models~\cite{Lesgourgues_etal_2009}.  References~\cite{Carlson_White_Padmanabhan_2009,Upadhye_2014} tested Time-RG for a wide range of cosmological models by comparison to N-body simulations, and Ref.~\cite{Upadhye_2016} extended it to redshift space through the approach of Ref.~\cite{Taruya_etal_2010}.  At $z=1/2$ and scales relevant to current data, Time-RG was confirmed accurate to $<5\%$ in the monopole and $<10\%$ in the quadrupole for a range of massive neutrino models~\cite{Upadhye_2016}.  Reference~\cite{Beutler_2014} showed, and we confirm in Sec.~\ref{subsec:galaxies_as_biased_tracers}, that these remaining errors are absorbed into the scale-dependent bias parameters, resulting in a bias-marginalized power spectrum accurate to $\approx 1\%$.

N-body simulations containing light neutrinos $m_\nu \lesssim 1$~eV are computationally daunting. Their large velocity dispersions mean that simulations must follow the full six-dimensional phase space of neutrinos, rather than only their positions as for CDM.
Simulations including massive neutrino particles can capture the imprint of neutrinos on the CDM power, as well as the cross-correlation between CDM and neutrinos~\cite{Viel_Haehnelt_Springel_2010}, but that reference notes that the neutrino power spectrum itself may be affected by shot noise at scales of interest to modern galaxy surveys. Reference~\cite{Bird_Viel_Haehnelt_2012} develops a power spectrum fitting function based on such simulations, though it has not yet been extended to redshift space and is therefore not used here.  Progress continues in the field of neutrino simulations, with new techniques under development~\cite{Banerjee_Dalal_2016}.
From the perturbation theory side, Reference~\cite{Fuhrer_2015} considers non-linear neutrino perturbations beyond the fluid approximation.  Integrating their approach into the redshift-space Time-RG of Ref.~\cite{Upadhye_2016} using the FFT techniques of Ref.~\cite{McEwen_2016} is a promising avenue for improving the accuracy of the redshift-space power spectrum, particularly if it can be verified by comparison to next-generation massive neutrino N-body simulations.

For the moment, we adopt the much simpler linearized neutrino approximation of Refs.~\cite{Saito_2008,Agarwal_2011}, which approximates the coordinate-space matter power spectrum as $P(k) = \fcb^2 \Pdd + 2 \fcb \fnu \sqrt{\Pdd \Pnu} + \fnu^2 \Pnu$ using the linear neutrino power spectrum $\Pnu(k)$.  In particular, we approximate the scale-dependent correlation coefficient $P_{\delta_\mathrm{CB}\delta_\nu} / \sqrt{P_{\delta_\mathrm{CB}\delta_\mathrm{CB}} P_{\delta_\nu \delta_\nu}}$ as unity.  Generalizing to redshift space, we make the following ansatz for the  power spectrum $P_s(k,\mu)$ in massive neutrino models:
\begin{eqnarray}
  P_s
  &=&
  \Ffog \!\!\cdot\!\!
  \left[
    \fcb^2(\Pdd \!+\! 2\mu^2 \! \Pdt \!+\! \mu^4 \! \Ptt
    \!+\! \Pbis \!+\! \Ptri)
    \!+\! \Peffnu\!
    \right]\qquad
\end{eqnarray}
where
\begin{eqnarray}
  \Peffnu
  &=&
  P_\mathrm{eff}^{(\nu,0)}
  + \mu^2 P_\mathrm{eff}^{(\nu,2)}
  + \mu^4 P_\mathrm{eff}^{(\nu,4)}
  \label{e:Peffnu}
  \\
  P_\mathrm{eff}^{(\nu,0)}
  &=&
  \fnu^2 \Pnu + 2\fnu\fcb \sqrt{\Pdd \Pnu}
  \\
  P_\mathrm{eff}^{(\nu,2)}
  &=&
  2\fnu^2 {\tilde f} \Pnu
  + 2 \fnu \fcb \sqrt{ \Ptt \Pnu }
  \nonumber\\
  &~&
  + 2 \fnu \fcb {\tilde f} \sqrt{\Pdd \Pnu}
  \\
  P_\mathrm{eff}^{(\nu,4)}
  &=&
  \fnu^2 {\tilde f}^2 \Pnu
  + 2 \fnu \fcb {\tilde f} \sqrt{\Ptt \Pnu}  
  \\
  \tilde{f}
  &=&
  \partial \log(\delta_{\nu,\mathrm{lin}}) / \partial \log(a).
\end{eqnarray}
Here, the quantities $\delta$ and $\theta$ with no subscripts refer to the non-linear $\delta_\mathrm{CB}$ and $\theta_\mathrm{CB}$, respectively, a convention which we use henceforth.  The above ansatz approximates the massive neutrinos as a linear fluid with velocity divergence $\theta_\nu \approx {\tilde f}\delta_\nu$. We will briefly test this ansatz in Sec.~\ref{sec:results_and_discussion}.

\subsection{Galaxies as biased tracers}
\label{subsec:galaxies_as_biased_tracers}

Thus far we have considered the density and velocity fields of the matter in the universe.  What galaxy surveys actually observe is the number density of a sample of galaxies selected in a precise way.  The mismatch, or ``bias,'' between the galaxies and the matter, is typically characterized in one of two ways.  Top-down bias approaches directly estimate large-scale statistical observables such as the power spectrum by modeling the galaxy density $\delta_\mathrm{g}$ as a function of the cosmological perturbations.  Bottom-up approaches model the way that galaxies populate individual dark matter halos based on the properties of those halos.  Given the distribution of these galaxies, a galaxy power spectrum can then be computed.  Here we discuss simple examples of both approaches, and argue that they give very similar results over the range of scales probed by modern galaxy surveys.

An elegant approach to the top-down modeling of galaxy bias was provided by McDonald and Roy (MR) in Ref.~\cite{McDonald_Roy_2009}.  We will use a variant of this MR bias model in Sec.~\ref{sec:results_and_discussion}, so we summarize it here and defer more extensive details to Appendix~\ref{sec:scale-dependent_bias}.  MR argues that, since the power spectra are constructed from the scalar variables $\delta$ and $\theta$, the most general galaxy density field at a given order in perturbation theory is a linear combination of all scalar variables up to that order.  The galaxy power spectra appropriate for the one-loop perturbation theory used here are
\begin{eqnarray}
\Pgdd\!\!\!
&=&
\bd^2 \Pdd
+ 2 \bd \bdd P_{\delta\delta^2}
+ 2 \bd \bss P_{\delta s^2}
+ \bdd^2 P_{\delta^2\delta^2}
\nonumber \\
&~&
+ 2 \bdd \bss P_{\delta^2s^2}
+ \bss^2 P_{s^2s^2}
+ 2 \bd \btnl P_{3\mathrm{nl}}
+ N\quad
\\
\Pgdt\!\!\!
&=&
\bd \bv \Pdt
\!+\! \bdd P_{\theta\delta^2}
\!+\! \bss P_{\theta s^2}
\!+\! \btnl f P_{3\mathrm{nl}}
\qquad
\\
\Pgtt\!\!\!
&=&
\Ptt,
\end{eqnarray}
with the integrals $P_{\delta\delta^2}(k)$, $P_{\theta\delta^2}(k)$, $P_{\delta s^2}(k)$, $P_{\theta s^2}(k)$, $P_{\delta^2\delta^2}(k)$, $P_{\delta^2s^2}(k)$, $P_{s^2s^2}(k)$, and $P_{3\mathrm{nl}}(k)$ given in Eqs.~(\ref{e:Pdd2}-\ref{e:P3nl}) of Appendix~\ref{sec:scale-dependent_bias}. Thus there are five bias parameters: $\bd$, $\bdd$, $\bss$, $\btnl$, and the shot noise $N$.

As a simplification we can work in a restricted subset of this five-dimensional bias parameter space.  Thus far, only symmetry arguments have been invoked to characterize the bias.  Working in a local Lagrangian evolution model, Ref.~\cite{Saito_etal_2014} finds
\begin{eqnarray}
  \bss  &=& -\frac{4}{7} (\bd - 1)
  \label{e:bss_const}
  \\
  \btnl &=& \frac{32}{315} (\bd - 1).
  \label{e:btnl_const}
\end{eqnarray}
Thus we consider two bias models:
\begin{enumerate}
\item MR($3$-param), a $3$-parameter model in which $\bd$, $\bdd$, and $N$ are allowed to vary, with $\bss$ and $\btnl$ fixed as in Eqs.~(\ref{e:bss_const},\ref{e:btnl_const}); and
\item MR($5$-param), varying $\bd$, $\bdd$, $\bss$, $\btnl$, and $N$.
\end{enumerate}

Bottom-up models such as Halo Occupation Distributions (HOD)~\cite{Zheng_etal_2005} instead model the average numbers of galaxies within dark matter halos, from which the galaxy power spectrum may subsequently be calculated. The key simplifying assumption is that these galaxy properties depend only on the halo mass, rather than its environment or history.  For example, Ref.~\cite{Zheng_etal_2009} models the expected number of central and satellite galaxies for a halo of mass $M$ as, respectively,
\begin{eqnarray}
\left<n_\mathrm{cen}\right>
&=&
\frac{1}{2}\mathrm{erfc} 
\left(\frac{\log(M_\mathrm{cut}/M)}{\sqrt{2}\sigma}\right),
\label{e:zheng_hod0}
\\
\left<n_\mathrm{sat}\right>
&=&
\left(\frac{M - \kappa M_\mathrm{cut}}{M_1}\right)^\alpha,
\label{e:zheng_hod1}
\end{eqnarray}
where $M_\mathrm{cut}$, $M_1$, $\alpha$, $\kappa$, and $\sigma$ are free parameters.  A given halo may have at most one central galaxy, and halos with central galaxies may also have satellites. The halos themselves may be found from N-body simulations, with halo particles chosen at random to be labeled satellites.

Running an N-body simulation and computing a power spectrum from a large number of simulated galaxies are numerically expensive.  However, Ref.~\cite{Kwan_etal_2015} has constructed an emulator, a statistical interpolation of the HOD power spectrum associated with Eqs.~(\ref{e:zheng_hod0},\ref{e:zheng_hod1}), for a cosmology with parameters $\ns=0.963$, $\sigma_8=0.8$, $h=0.71$,  $\omegam=0.1335$, $\omegab=0.02258$, and $\omeganu=0$.  This emulator allows for the rapid computation of galaxy power spectra at the percent level as a function of the five HOD parameters, though further work is required to generalize it beyond this particular cosmological model.

As a way of testing the accuracy of the MR model used in this article, we compare it against $1000$ randomly-generated HOD power spectra using the emulator.  We choose the HOD parameters from uniform random distributions within the bounds covered by the emulator, ranges chosen to approximate BOSS CMASS galaxies: $10^{12.85} < M_\mathrm{cut}/M_\odot < 10^{13.85}$, $10^{13.3} < M_1 /M_\odot < 10^{14.3}$, $0.5 < \alpha < 1.5$, $0.5 < \kappa < 1.5$, and $0.5 < \sigma < 1.2$.  For each HOD parameter set $N$, we minimize 
\begin{equation}
\chi^2_N(\vec b)
=
\sum_i \frac{\left[ \Pgdd(k_i,\mu=0;\vec b) - P_N(k_i) \right]^2}{P_N(k_i)^2}
\end{equation}
in $100$ logarithmically-spaced wave number bins between $0.005~h/\mathrm{Mpc} \leq k \leq 0.2~h/\mathrm{Mpc}$.  The power spectra are evaluated at $z=0.57$ characteristic of the SDSS BOSS data~\cite{Beutler_2014}, and at $\mu=0$ since redshift-space distortions are not emulated.  

\begin{figure*}[tb]

  \raisebox{55mm}[0pt][0pt]{\makebox[0pt][s]{~~{\bf (a)}}}
  \includegraphics[width=3.4in]{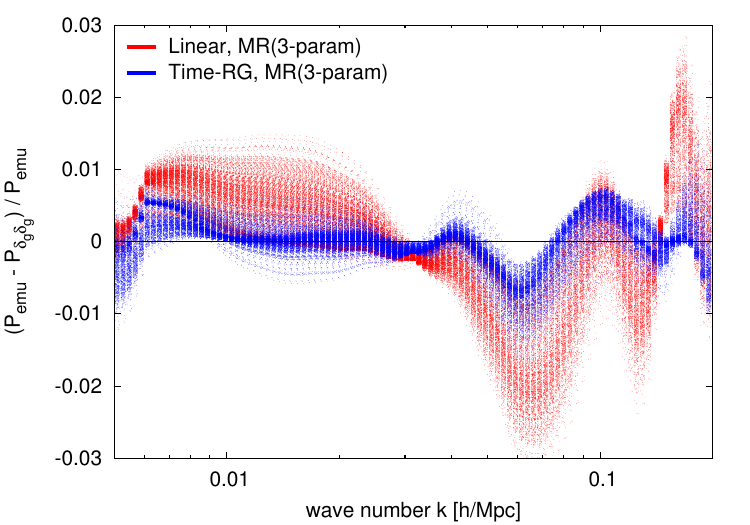}
  \raisebox{55mm}[0pt][0pt]{\makebox[0pt][s]{~~{\bf (b)}}}
  \includegraphics[width=3.4in]{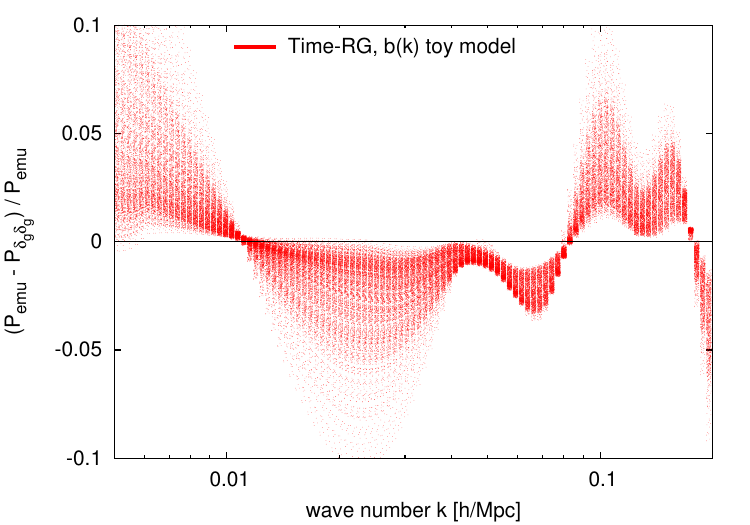}%

  \raisebox{55mm}[0pt][0pt]{\makebox[0pt][s]{~~{\bf (c)}}}
  \includegraphics[width=3.4in]{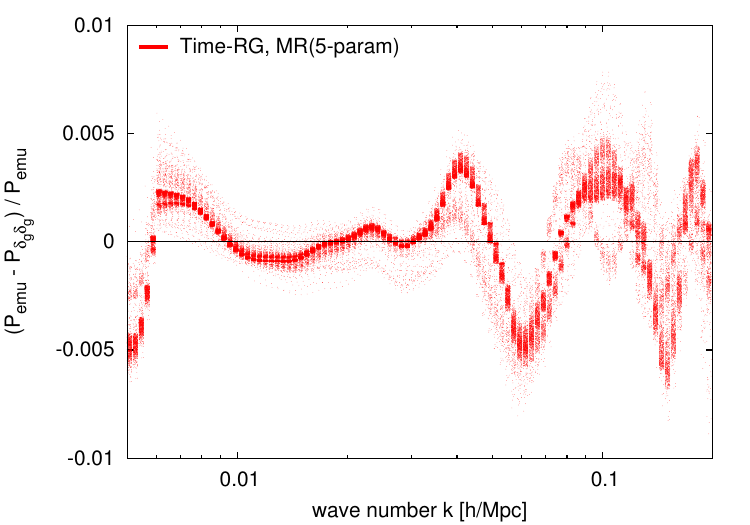}
  \raisebox{55mm}[0pt][0pt]{\makebox[0pt][s]{~~{\bf (d)}}}
  \includegraphics[width=3.4in]{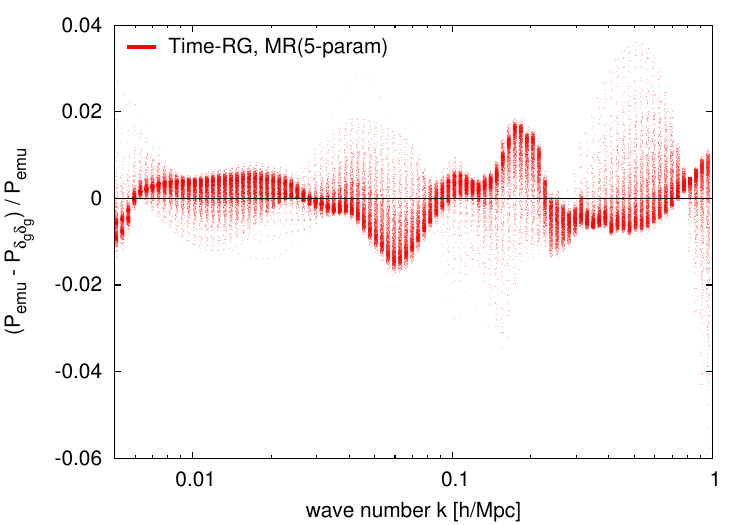}%

  \caption{
    Perturbation theory with top-town bias models compared with HOD models.
    (a)~MR with $3$ parameters, using linear or Time-RG perturbation theory
    for $\Pdd$.  Switching from linear theory to 
    Time-RG  reduces errors by factors of $2$-$3$.
    (b)~Time-RG with a simple $3$-parameter bias model 
    $b(k) = (b_0 + b_1 k)/(1 + b_2 k^2)$, which has substantially larger
    errors than the MR model with $3$ parameters.
    (c)~Time-RG with $5$-parameter MR model. Systematic errors are 
    approximately as large as with the $3$-parameter MR model.
    (d)~Time-RG with $5$-parameter MR model fit in the range
    $0.005~h/\mathrm{Mpc} < k < 1~h/\mathrm{Mpc}$.
    \label{f:bias_tests}
  }
\end{figure*}

Figure~\ref{f:bias_tests} shows $1 - \Pgdd(k_i,0;\vec b) / P_N(k_i)$, the binned residuals at $\mu=0$, for different bias models and perturbation theories. Time-RG plus the $3$-parameter MR model, the combination used in most of this article, is shown in Fig.~\ref{f:bias_tests}(a), where its errors improve upon those of linear theory by factors of $2$-$3$.  The resulting residuals are less than $1\%$, consistent with emulator errors.  This error estimate is the key result of this section.  

The MR model is substantially more accurate than a simple toy bias model $b(k) = (b_0 + b_1 k)/(1 + b_2 k^2)$, as seen in Fig.~\ref{f:bias_tests}(b), even though the two fit the same number of parameters.  Fig.~\ref{f:bias_tests}(c) shows that the $5$-parameter MR model does not substantially improve upon the $3$-parameter model over the range of scales probed by BOSS; though the scatter among points is smaller, a $\approx 0.5\%$ systematic error remains.  Finally, Fig.~\ref{f:bias_tests}(d) extends the fitting range to $k=1~h/$Mpc and shows that the full $5$-parameter MR model remains accurate at the $2\%-3\%$ level over nearly the entire range.

\begin{figure}[tbp]
  \includegraphics[width=3.4in]{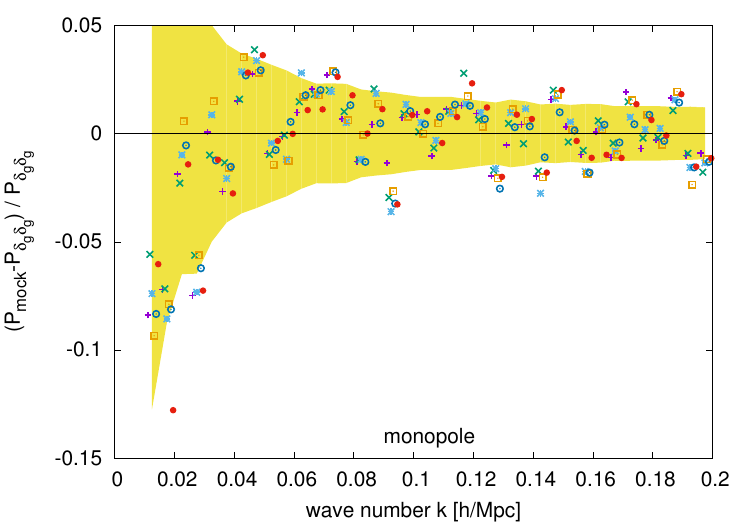}
  \caption{
    Time-RG perturbation theory with the MR($3$-param) bias model compared with
    mock HOD power spectrum monopoles for a model with $\omeganu=0.0009$
    ($\sum m_\nu = 84$~meV).  Yellow bands show BOSS DR11 error bars.
    \label{f:rsd_hod_mock_mono}
  }
\end{figure}

\begin{figure}[tbp]
  \includegraphics[width=3.4in]{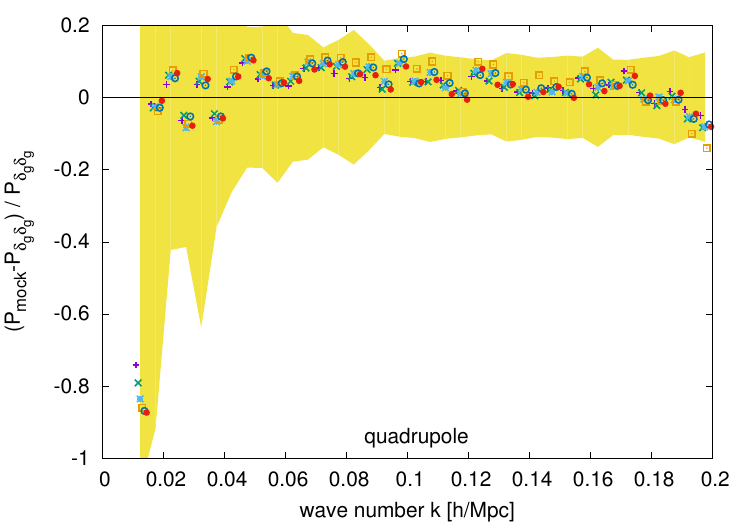}
  \caption{
    Same as Fig.~\ref{f:rsd_hod_mock_mono} but for the quadrupoles.
    \label{f:rsd_hod_mock_quad}
  }
\end{figure}

Next, although an emulator such as that of Ref.~\cite{Kwan_etal_2015} is not available in redshift space, we have obtained six mock HOD power spectra from the same simulations as that reference, for a massive-neutrino cosmology with parameters $\ns=1.0005$, $\sigma_8=0.8812$, $h=0.8019$, $\omegac=0.1145$, $\omegab=0.0215$, $\omeganu=0.000905$, $w_0=-0.7282$, and $w_a=-1.6927$.  Figures~\ref{f:rsd_hod_mock_mono} and \ref{f:rsd_hod_mock_quad} compare these mocks to Time-RG perturbation theory with the best-fitting MR($3$-param) bias parameters.  The fit was computed using the BOSS DR11 covariance matrix, whose error bars are somewhat larger than the scatter among simulation points, leading to a $\chi^2/\mathrm{d.o.f.}$ of $0.6$.  Thus while there may be residual differences between top-down and bottom-up bias models, especially for the quadrupole, these are subdominant to the current experimental uncertainties.  Future data sets will likely require more sophisticated modeling of the redshift-space galaxy power spectrum.

This broad agreement between the top-down McDonald-Roy bias model and the bottom-up HOD approach over the range of scales $k \lesssim 0.2~h/$Mpc gives us confidence that the MR models can approximate the galaxy power spectrum monopole at the $0.5\% - 1\%$ error level over the entire range accessible to modern galaxy surveys, and the quadrupole to well within current measurement errors.  For the remainder of this work we apply the MR bias models, with MR($3$-param) used unless otherwise noted.  Further details of our bias implementation may be found in Appendix~\ref{sec:scale-dependent_bias}.

Finally we briefly comment on neutrinos in bias modeling.  Following Ref.~\cite{Villaescusa-Navarro_2014} we have applied bias corrections to the CDM+baryon power spectrum, rather than the total matter power spectrum.  One may ask, however, how sensitive the galaxy density is to the underlying distribution of neutrinos.  Since galaxies form from baryons, which are correlated with the CDM, and since the neutrino halos around CDM halos are diffuse~\cite{LoVerde_Zaldarriaga_2014}, the galaxy density may be  weakly dependent on $\Pnu$. Thus we define a neutrino bias $\bnu$ by
\begin{eqnarray}
  \Peffnu(k,\mu,\bnu)
  &=&
  P_\mathrm{eff}^{(\nu,0)}
  + \mu^2 P_\mathrm{eff}^{(\nu,2)}
  + \mu^4 P_\mathrm{eff}^{(\nu,4)}
  \label{e:Peffnu_bias}
  \\
  P_\mathrm{eff}^{(\nu,0)}(k,\bnu)
  &=&
  \fnu^2 \bnu^2 \Pnu + 2\fnu\fcb \bnu \sqrt{\Pdd \Pnu}
  \\
  P_\mathrm{eff}^{(\nu,2)}(k,\bnu)
  &=&
  2\fnu^2 {\tilde f} \bnu \Pnu
  + 2 \fnu \fcb \bnu \sqrt{ \Ptt \Pnu }
  \nonumber\\
  &~&
  + 2 \fnu \fcb {\tilde f} \sqrt{ \Pdd \Pnu }
  \\
  P_\mathrm{eff}^{(\nu,4)}(k,\bnu)
  &=&
  \fnu^2 {\tilde f}^2 \Pnu
  + 2 \fnu \fcb {\tilde f} \sqrt{\Ptt \Pnu}.\quad
\end{eqnarray}
Allowing $\bnu \neq 0$ may also be seen as a test of our ansatz in Eq.~(\ref{e:Peffnu}) for the neutrino contribution to the redshift-space power spectrum.  A strong dependence of cosmological parameters on $\bnu$ would motivate a more careful consideration of this ansatz.  Section~\ref{sec:results_and_discussion} compares results for $\bnu=0$ to $\bnu=1$.  Unless otherwise noted, we set $\bnu = 0$ from now on.  Appendix~\ref{sec:scale-dependent_bias} shows that the galaxy power spectrum including scale-dependent bias and $\bnu$ can be expressed as
\begin{equation}
  \frac{P(k,\mu)}{\Ffog(\mu\sigma_v k f)}
    =
    \sum_{L=0}^{25} \mu^{n_L} \fcb^2\!B_L\!(\vec b) P_L
    \!+\! \sum_{n=0}^4 \mu^n \Peffnun
    \! + \! N \!
\end{equation}
with exponents $n_L$, bias polynomials $B_L(\vec b)$, and power spectrum components $P_L(k)$ given in Table~\ref{t:Pkmu_L}.

\section{Data sets}
\label{sec:data_sets}

The major goals of this work are to use the scale-dependent redshift-space galaxy power spectrum to constrain the sum of neutrino masses, and to investigate the impacts of different bias models and cosmological parameters on this constraint.  Since our focus here is the contribution of the galaxy redshift survey, we include data from the cosmic microwave background and the type IA supernova Hubble diagram in the simplest reasonable ways.  This section describes our treatment of each data set, as well as our joint analysis using a Monte Carlo Markov Chain (MCMC) procedure.

\subsection{Galaxy survey: BOSS DR11}
\label{subsec:galaxy_survey_boss_dr11}

The primary galaxy survey observable analyzed in this article is the redshift-space CMASS power spectrum of BOSS Data Release 11 (BOSS DR11), computed from the data in Ref.~\cite{Beutler_2014} and applied to neutrino masses in Ref.~\cite{Beutler_2014b}.  We focus on DR11 rather than the more recent release of Refs.~\cite{Alam_2016,Beutler_2017} because the publicly-released DR11 Fourier-space power spectrum data cover a larger range of scales, up to $k_\mathrm{max}=0.2~h/$Mpc, with a simpler bias structure.  We briefly discuss our choice at the end of this subsection, and we compare DR11 to DR12 constraints in the next section.

The CMASS galaxy sample of BOSS DR11 consists of massive, high-redshift galaxies, $0.43 \leq z < 0.7$, with biases $\bd \approx 2$, which are typically central rather than satellite galaxies~\cite{Ross_2012}.  Reference~\cite{Beutler_2014} computed the power spectrum, using $690,827$ galaxies observed over an area of $8498$ square degrees, by applying the estimator of Ref.~\cite{Yamamoto_2006}.  Data products provided by BOSS DR11 are:
\begin{enumerate}
\item the monopole and quadrupole of the measured redshift-space power
  spectrum $P(k,\mu)$, in $38$ $k$ bins of width $\Delta k = 0.005~h/$Mpc,
  covering the range $0.01~h/\mathrm{Mpc} \leq k \leq 0.2~h/\mathrm{Mpc}$;
\item window functions $w_{\ell,\ell'}(k,k')$ to be convolved with a model
  power spectrum before comparison with the data;
\item the covariance matrix ${\bf C}$ of the data.
\end{enumerate}

Our likelihood computation procedure is based on that of Ref.~\cite{Beutler_2014}.  Briefly, we use the data and covariance matrix to find the model-dependent likelihood
\begin{equation}
  -2 \log({\mathcal L})
  =
  \sum_{i,j} ({\mathbf C}^{-1})_{ij}
  (P_i^\mathrm{d}-P_i^\mathrm{t})(P_j^\mathrm{d}-P_j^\mathrm{t}),
\end{equation}
where $P_i^\mathrm{t}$ is the windowed model power spectrum in bin $i$ and $P_i^\mathrm{d}$ the data.  At each point in cosmological parameter space, we marginalize over the nuisance parameters, the velocity dispersion $\sigma_v$ and the biases $\vec b$.  We use the flat prior $0 < \sigma_v < 7$~Mpc/$h$ on $\sigma_v$ and open priors on $\vec b$.

Appendix~\ref{sec:boss_likelihood} shows that all bias-dependent terms can be factored out of $\log(\mathcal L)$, resulting in an fourth-order polynomial in $\vec b$.  This function is simple enough that we marginalize over $\vec b$ by direct numerical integration of $\mathcal L$ using the CUBA library of Ref.~\cite{Hahn_2005}.  We then integrate numerically over $\sigma_v$.  On a standard eight-processor computing node in the HTCondor cluster at the University of Wisconsin-Madison, a single likelihood evaluation at a point in cosmological parameter space takes $15-20$~sec.~for the MR($3$-param) bias model, which includes several seconds each to run \mbox{\Camb} and \redtime.

The more recent BOSS $2$-D Fourier-space power spectrum of Ref.~\cite{Beutler_2017} traded higher angular resolution, in the form of a measured hexadecapole, for worse spatial resolution, with a maximum wave number $k_\mathrm{max} = 0.15~h/$Mpc rather than $0.20~h/$Mpc.  It also had different galaxy populations in the northern and southern sky patches, which when combined with the three redshift bins, led to a sixfold increase in the number of bias parameters.  The neutrino mass constraint is especially sensitive to the shape of the power spectrum, hence the scale-dependent bias.  We will see in Sec.~\ref{subsec:massive_neutrinos_in_the_nuLCDM_model} that this smaller $k_\mathrm{max}$ and more complicated bias structure, in the context of our analysis based on open priors for $\vec b$, makes our DR12 constraints on $\sum m_\nu$ weaker than those of DR11. In the interests of providing the strongest possible constraints, we use DR11 data for all other analyses.

We note for completeness that the BOSS DR11 and DR12 galaxy data are publicly available, allowing for the independent calculation of the power spectrum to higher $k$.  However, thorough testing of these higher-$k$ data and the corresponding window functions is necessary before using them to constrain cosmology.  Such power spectrum calculation and testing are beyond the scope of the current article, and we restrict ourselves to the maximum wave numbers recommended by the authors, or smaller values.

\subsection{Cosmic Microwave Background: Planck}

Our treatment of the CMB data is straightforward, as we use the likelihood software provided by the Planck collaboration in Refs.~\cite{Adam_2016j,Aghanim_2016}.  Planck measured $C_\ell^{TT}$ over the range $2 \leq \ell \leq 2508$ as well as $C_\ell^{TE}$ and $C_\ell^{EE}$ over $2 \leq \ell \leq 1996$.  We use the {\tt bflike} likelihood of Ref.~\cite{Adam_2016j} to analyze $C_\ell^{TT}$, $C_\ell^{TE}$, and $C_\ell^{EE}$ for $2 \leq \ell \leq 29$.  The temperature and polarization power spectra for $\ell \geq 30$ are analyzed using the \plik-lite function of Ref.~\cite{Aghanim_2016}.  This function is marginalized over all nuisance parameters except an absolute calibration parameter $A_\mathrm{planck}$.  At each point in cosmological parameter space, we marginalize over this parameter in the recommended interval $0.9975 \leq A_\mathrm{planck} \leq 1.0025$.

The Planck CMB likelihood evaluation procedure used here for cosmological parameters $\vec c$ is as follows:
\begin{enumerate}
\item run \Camb~with a fiducial value of the scalar amplitude $A_s$ to
  find $\sigma_8$;
\item rescale $A_s$ to yield the desired $\sigma_8$, and rerun \Camb;
\item for a given $A_\mathrm{planck}$, use \plik-lite with the lensed $C_\ell$s
  to compute the likelihood ${\mathcal L}(\vec c, A_\mathrm{planck})$;
\item marginalize over $A_\mathrm{planck}$ by repeating the previous step over
  the range $0.9975 \leq A_\mathrm{planck} \leq 1.0025$.
\end{enumerate}

\subsection{Type Ia supernovae: JLA}

Since the CMB and galaxy survey power spectra are insufficient for constraining the evolution of the dark energy equation of state, our analysis also included Type Ia supernovae from the Joint Likelihood Analysis (JLA) of Ref.~\cite{Betoule_2014}.  We use the compressed likelihood of that reference, which the authors confirmed to match the mean values and uncertainties of the full likelihood to $0.018\sigma$ and $0.3\%$, respectively.  The likelihood $\mathcal L(\vec c, \mathcal M)$ is given in terms of cosmological parameters $\vec c$ and a nuisance parameter $\mathcal M$ by
\begin{eqnarray}
  -2\log(\mathcal L(\vec c, \mathcal M))
  &=&
  \sum_{i,j} ({\mathbf C}^{-1})_{ij} (r_i-\mathcal M) ( r_j - \mathcal M)\qquad
  \\
  r_i
  &=&
  \mu_i - 5 \log_{10}(D_\mathrm{L}(\vec c, z_i))
  \\
  D_\mathrm{L}(\vec c, z_i)
  &=&
  (1+z) H_0 \chi(\vec c, z_i)
\end{eqnarray}
where the binned magnitude parameters $\mu_i$ and the covariance matrix $\mathbf C$ are provided by Ref.~\cite{Betoule_2014}.  Marginalization over $\mathcal M$ is straightforward:
\begin{eqnarray}
  \log(\mathcal L(\vec c))
  &=&
  -\frac{1}{2}\chi_0^2 + \frac{1}{2} f_0 \mathcal M_0^2
  + \frac{1}{2} \log\left(\frac{2\pi}{f_0}\right)
  \\
  f_0
  &=&
  \sum_{i,j} ({\mathbf C}^{-1})_{ij}
  \\
  \mathcal M_0
  &=&
  \frac{1}{f_0}
  \sum_{i,j} ({\mathbf C}^{-1})_{ij} r_j
  \\
  \chi_0^2
  &=&
  \sum_{i,j} ({\mathbf C}^{-1})_{ij} r_i r_j.
\end{eqnarray}

\subsection{Monte Carlo Markov Chain analysis}

In this article we consider nine cosmological parameters: the scalar spectral index $n_s$; the power spectrum amplitude $\sigma_8$ in $8$~Mpc$/h$ spheres; the dimensionless Hubble constant $h = H_0 / (100~\mathrm{km / sec / Mpc})$; the CDM density $\omegac = \Omegaco h^2$; the baryon density $\omegab = \Omegabo h^2$; the neutrino density $\omega_\nu = \Omega_{\nu 0} h^2$; the optical depth $\tau$ to the surface of last scattering; and the dark energy parameters $w_0$ and $w_a$ specifying the time-dependent equation of state $w(z) = P_\mathrm{de} / \rho_\mathrm{de} = w_0 + w_a z / (1+z)$.  We are specifically interested in the following subsets fixing some of the parameters:
\begin{itemize}
\item $\Lambda$CDM: $\omeganu = 0.0006$, $w_0 = -1$, $w_a = 0$;
\item $\nu\Lambda$CDM: $w_0 = -1$, $w_a = 0$;
\item $w$CDM: $\omeganu = 0.0006$;
\item $\nu w$CDM: all $9$ parameters allowed to vary.
\end{itemize}

Our Markov chains cover the same parameter space with a different set of parameters.  Rather than $\sigma_8$ we use $1 + \log(\sigma_8^2)$ for ease of comparison with \CosmoMC~\cite{Lewis_Bridle_2002}.  Since $h$ is covariant with several other cosmological parameters, we instead use $100$ times the characteristic angular scale of the CMB acoustic oscillations,
\begin{eqnarray}
  \theta_{100}
  &=&
  100 \times r_s / \chi(z_\mathrm{CMB})
  \\
  r_s
  &=&
  \frac{55.234 h~\mathrm{Mpc}/h}
  {(\omegab + \omegac)^{0.2538} \omegab^{0.1278} (1+\omeganu)^{0.3794}}
\end{eqnarray}
where $z_\mathrm{CMB}$ is the redshift of the surface of last scattering; we use the fitting function of Ref.~\cite{Anderson_2014} for the comoving sound horizon $r_s$ at decoupling.  For a general dark energy $w(z)$, conversion from $\theta_{100}$ to $h$ is done iteratively starting from a guessed value of $h$. In the special case of a cosmological constant, $w(z) = -1$, an approximate form of $\theta_{100}$ makes this conversion simpler:
\begin{eqnarray}
  \theta_{100}
  &=&
  \frac{F_{100}(\omegac,\omegab,\omeganu)~y_0^{-1/2}}{I(y_\mathrm{CMB}) - I(y_0)}
  \\
  F_{100}
  &=&
  \frac{1.8424~\omegam^{1/2}}{(\omegab + \omegac)^{0.2538} \omegab^{0.1278}
    (1+\omeganu)^{0.3794}}
  \\
  I(y)
  &=&
  \int_0^y \frac{dy'}{\sqrt{1+(y')^3}}
  \\
  y_0
  &=&
  (\Omegamo / \Omega_\Lambda)^{1/3}
  ~\mathrm{and}~y(z) = (1+z)y_0.
\end{eqnarray}
The integral $I(y)$ can be approximated in the low-$y$ and high-$y$ limits:
\begin{eqnarray}
  I(y_0)
  &\approx&
  y_0 - y_0^4/8~\mathrm{for~low}~y_0
  \\
  I(y_\mathrm{CMB})
  &\approx&
  2.8043 - 2 / \sqrt{y_\mathrm{CMB}}
  ~\mathrm{for}~y_\mathrm{CMB} \gg 1.
\end{eqnarray}
The high-$y$ approximation is accurate to $0.01\%$ for $y > 9$.  For $y < 1.3$, corresponding to $\Omegamo < 0.69$, the low-$y$ approximation is accurate at the $10\%$ level, which is sufficient for breaking parameter degeneracies. Given $\theta_{100}$, $\omegac$, $\omegab$, and $\omeganu$ (hence $F_{100}$) at a point in cosmological parameter space, we find $h$ by guessing $y_0$, computing $I(y_\mathrm{CMB})-I(y_0)$, and refining our guess to $y_0 = F_{100}^2 / \theta_{100}^2 / [I(y_\mathrm{CMB})-I(y_0)]^2$.  Once this iteration has converged, $h = \omegam^{1/2} \sqrt{1 + y_0^{-3}}$.

\begin{table}[tb]
  \tabcolsep=0.15cm
  \begin{footnotesize}
    \begin{tabular}{c|c|c|c|c|c|c|c}
      $\ns$ & $\sigma_8$ & $h$ & $\omegac$ & $\omegab$ & $\omeganu$ & $w_0$,$w_a$ & $\tau$ \\
      \hline
      $>0$  & $>0$   & $[0.2,1]$ & $>0$   & $>0.001$ & $\geq 0$ & $w_0+w_a\leq 0$ & $>0.01$
    \end{tabular}
  \end{footnotesize}
  \caption{
    The prior probability distribution is uniform in the parameters $n_s$,
    $1+\log(\sigma_8^2)$, $\theta_{100}$, $\omegac$, $\omegab$, $\omeganu$,
    $w_0$, $w_a$, and $\tau$, with the above bounds.  As noted in
    Sec.~\ref{subsec:galaxy_survey_boss_dr11}, we
    also impose  $0 < \sigma_v < 7$~Mpc/$h$ on the velocity dispersion
    length, and open priors on the biases $\vec b$.
    \label{t:priors}
  }
\end{table}

We impose a minimal set of prior constraints on the cosmological parameter space.  Since $w(z) \rightarrow w_0 + w_a$ as $z \rightarrow \infty$, a positive value of $w_0 + w_a$ would imply that the dark energy density would grow faster than the matter density at high redshift.  Thus we require $w_0 + w_a \leq 0$.  We require that $n_s$, $h$, $\sigma_8$, $\omegac$, $\omegab$, $\omeganu$, and $\tau$ all be non-negative.  In order to use \mbox{\Camb} we require: $0.2 \leq h \leq 1$; $\omegab \geq 0.001$; and $\tau > 0.01$.  We also assume a spatially flat universe, $\Omega_K = 0$.  Within the allowed intervals, our priors are uniform in the chain parameters $n_s$, $1+\log(\sigma_8^2)$, $\theta_{100}$, $\omegac$, $\omegab$, $\omeganu$, $w_0$, $w_a$, and $\tau$. The priors used are summarized in Table~\ref{t:priors}.

Each chain is initialized by choosing the parameters allowed to vary from uniform random distributions over the intervals $0.93 \leq n_s \leq 0.99$, $0.25 \leq 1 + \log(\sigma_8^2) \leq 0.85$, $1.047 \leq \theta_{100} \leq 1.053$, $0.1133 \leq \omegac \leq 0.1265$, $0.0212 \leq \omegab \leq 0.023$, $0.0015 \leq \omeganu \leq 0.0045$, $-1.3 \leq w_0 \leq -0.7$, $-1.5 \leq w_a \leq 1.3$, and $0.035 \leq \tau \leq 0.125$. 

Our Markov chain analysis uses the standard Metropolis-Hastings procedure.  At each step in the chain, a new point is proposed using a fixed, symmetric proposal function.  Points can be chosen more efficiently by accounting for the covariances among the chain parameters.  If the chain parameter covariance matrix ${\mathbf C}$ has normalized eigenvectors $\{\hat x^{(i)}\}$ with corresponding eigenvalues $\{\lambda^{(i)}\}$, then the most efficient proposed steps would be linear combinations $\sum_i a^{(i)} \sqrt{\lambda^{(i)}} \hat x^{(i)}$ added to the current point.  Since we would like the chain to be able to take large steps $a^{(i)} \sim 1$ as well as smaller steps $a^{(i)} \sim 0.01 - 0.1$ in order to navigate out of narrow valleys in the likelihood surface, we choose the $a^{(i)}$ as follows.  First we select a random integer $0 \leq r \leq 9$ and define $\sigma_r = 10^{2r/9 - 2}$.  Then, for $I$ cosmological parameters, we choose $\{a^{(i)}\}$ randomly from an $I$-dimensional Gaussian of width $\sigma_r$.  Following the standard MCMC procedure, we accept the step from point $\vec c$ to ${\vec c}\,' = \vec c + \sum_i a^{(i)} \sqrt{\lambda^{(i)}} \hat x^{(i)}$ if its likelihood is greater, ${\mathcal L}(\vec c\,') > {\mathcal L}(\vec c)$.  Otherwise we accept it with probability ${\mathcal L}(\vec c\,') / {\mathcal L}(\vec c)$.  Our parameter covariance matrices ${\mathbf C}$ are the ``{\tt base\_BAORSD\_TTTEEE\_lowTEB\_plik.covmat,}'' ``{\tt base\_mnu\_BAORSD\_TTTEEE\_lowTEB\_plik.covmat},'' and ``{\tt base\_w\_wa\_BAO\_HST\_JLA\_TTTEEE\_lowTEB\_plik.covmat}'' matrices provided with the \mbox{\CosmoMC} code.

For each combination of cosmological model, bias parameterization, and data combination, we run a set of five Markov chains.  Convergence of each set is assessed using the test of Brooks and Gelman in Refs.~\cite{Gelman_Rubin_1992,Brooks_Gelman_1997}.  The latter reference defines a potential scale reduction factor $R_c^{1/2}$ which approaches $1$ from above as the variance of means within each set becomes much smaller than the mean of variances.  The authors recommend $R_c^{1/2} < 1.2$ as a standard of convergence.  We use the more stringent standard $R_c^{1/2} < 1.1$ for all but the BOSS DR12 analyses.

\section{Results and discussion}
\label{sec:results_and_discussion}

\subsection{Vanilla $\Lambda$CDM model}
\label{subsec:vanilla_LCDM_model}


\begin{table*}[tb]
  \tabcolsep=0.02cm
  \begin{footnotesize}
    \begin{tabular}{r||c|c|c|c|c}
      &
      $\Lambda$CDM, MR(3), PB
      &
      $\nu\Lambda$CDM, MR(3), PB
      &
      $\nu\Lambda$CDM, MR(3), PBJ
      &
      $w$CDM, MR(3), PBJ
      &
      $\nu w$CDM, MR(3), PBJ
      \\
      
      \hline
\hline
$n_s$
&
 $0.9625 \begin{array}{ll} {}_{+0.0041} & {}_{+0.0076} \\ {}^{-0.0039} & {}^{-0.0081} \end{array}$
&
 $0.963 \begin{array}{ll} {}_{+0.0041} & {}_{+0.0076} \\ {}^{-0.0042} & {}^{-0.0082} \end{array}$
&
 $0.9627 \begin{array}{ll} {}_{+0.0039} & {}_{+0.008} \\ {}^{-0.004} & {}^{-0.008} \end{array}$
&
 $0.9622 \begin{array}{ll} {}_{+0.0041} & {}_{+0.0085} \\ {}^{-0.0044} & {}^{-0.0084} \end{array}$
&
 $0.9613 \begin{array}{ll} {}_{+0.0044} & {}_{+0.0091} \\ {}^{-0.0045} & {}^{-0.009} \end{array}$
\\
\hline
$\sigma_8$
&
 $0.820 \begin{array}{ll} {}_{+0.013} & {}_{+0.025} \\ {}^{-0.013} & {}^{-0.025} \end{array}$
&
 $0.812 \begin{array}{ll} {}_{+0.014} & {}_{+0.028} \\ {}^{-0.014} & {}^{-0.028} \end{array}$
&
 $0.815 \begin{array}{ll} {}_{+0.016} & {}_{+0.029} \\ {}^{-0.014} & {}^{-0.03} \end{array}$
&
 $0.825 \begin{array}{ll} {}_{+0.015} & {}_{+0.028} \\ {}^{-0.013} & {}^{-0.029} \end{array}$
&
 $0.799 \begin{array}{ll} {}_{+0.024} & {}_{+0.044} \\ {}^{-0.024} & {}^{-0.044} \end{array}$
\\
\hline
$h$
&
 $0.6746 \begin{array}{ll} {}_{+0.0048} & {}_{+0.0091} \\ {}^{-0.0048} & {}^{-0.0094} \end{array}$
&
 $0.6722 \begin{array}{ll} {}_{+0.0063} & {}_{+0.011} \\ {}^{-0.0053} & {}^{-0.012} \end{array}$
&
 $0.6732 \begin{array}{ll} {}_{+0.0058} & {}_{+0.011} \\ {}^{-0.0049} & {}^{-0.011} \end{array}$
&
 $0.6788 \begin{array}{ll} {}_{+0.0074} & {}_{+0.015} \\ {}^{-0.0076} & {}^{-0.015} \end{array}$
&
 $0.6753 \begin{array}{ll} {}_{+0.0073} & {}_{+0.015} \\ {}^{-0.0077} & {}^{-0.015} \end{array}$
\\
\hline
$\omega_\mathrm{c}$
&
 $0.1193 \begin{array}{ll} {}_{+0.001} & {}_{+0.0021} \\ {}^{-0.0011} & {}^{-0.0021} \end{array}$
&
 $0.1191 \begin{array}{ll} {}_{+0.0011} & {}_{+0.0021} \\ {}^{-0.0011} & {}^{-0.0021} \end{array}$
&
 $0.1193 \begin{array}{ll} {}_{+0.0012} & {}_{+0.0022} \\ {}^{-0.0011} & {}^{-0.0022} \end{array}$
&
 $0.1194 \begin{array}{ll} {}_{+0.0013} & {}_{+0.0025} \\ {}^{-0.0014} & {}^{-0.0026} \end{array}$
&
 $0.1197 \begin{array}{ll} {}_{+0.0013} & {}_{+0.0027} \\ {}^{-0.0014} & {}^{-0.0026} \end{array}$
\\
\hline
$\omega_\mathrm{b}$
&
 $0.02222 \begin{array}{ll} {}_{+0.00014} & {}_{+0.00027} \\ {}^{-0.00013} & {}^{-0.00027} \end{array}$
&
 $0.02223 \begin{array}{ll} {}_{+0.00013} & {}_{+0.00027} \\ {}^{-0.00014} & {}^{-0.00026} \end{array}$
&
 $0.02223 \begin{array}{ll} {}_{+0.00014} & {}_{+0.00027} \\ {}^{-0.00014} & {}^{-0.00027} \end{array}$
&
 $0.02222 \begin{array}{ll} {}_{+0.00014} & {}_{+0.00028} \\ {}^{-0.00014} & {}^{-0.00029} \end{array}$
&
 $0.02217 \begin{array}{ll} {}_{+0.00016} & {}_{+0.00029} \\ {}^{-0.00015} & {}^{-0.00031} \end{array}$
\\
\hline
$\omega_\nu$
&
&
 $0.00098 \begin{array}{ll} {}_{+0.00044} & {}_{+0.00099} \\ {}^{-0.00071} & {}^{-0.00098} \end{array}$
&
 $0.00078 \begin{array}{ll} {}_{+0.00016} & {}_{+0.0011} \\ {}^{-0.00078} & {}^{-0.00078} \end{array}$
&
&
 $0.0029 \begin{array}{ll} {}_{+0.0018} & {}_{+0.0029} \\ {}^{-0.0029} & {}^{-0.0029} \end{array}$
\\
&
&
 ($<0.00197$ to $95\%$ CL)
&
 ($<0.00192$ to $95\%$ CL)
&
&
 ($<0.0058$ to $95\%$ CL)
\\
\hline
$w_0$
&
&
&
&
 $-0.87 \begin{array}{ll} {}_{+0.15} & {}_{+0.32} \\ {}^{-0.16} & {}^{-0.29} \end{array}$
&
 $-0.88 \begin{array}{ll} {}_{+0.15} & {}_{+0.32} \\ {}^{-0.17} & {}^{-0.33} \end{array}$
\\
\hline
$w_a$
&
&
&
&
 $-0.61 \begin{array}{ll} {}_{+0.76} & {}_{+1.2} \\ {}^{-0.61} & {}^{-1.3} \end{array}$
&
 $-0.90 \begin{array}{ll} {}_{+0.94} & {}_{+1.5} \\ {}^{-0.6} & {}^{-1.6} \end{array}$
\\
\hline
$\tau$
&
 $0.0672 \begin{array}{ll} {}_{+0.016} & {}_{+0.031} \\ {}^{-0.015} & {}^{-0.033} \end{array}$
&
 $0.0702 \begin{array}{ll} {}_{+0.017} & {}_{+0.034} \\ {}^{-0.017} & {}^{-0.033} \end{array}$
&
 $0.0676 \begin{array}{ll} {}_{+0.016} & {}_{+0.034} \\ {}^{-0.018} & {}^{-0.033} \end{array}$
&
 $0.0648 \begin{array}{ll} {}_{+0.017} & {}_{+0.036} \\ {}^{-0.016} & {}^{-0.032} \end{array}$
&
 $0.0728 \begin{array}{ll} {}_{+0.019} & {}_{+0.035} \\ {}^{-0.018} & {}^{-0.036} \end{array}$
\\
\hline
$\Omega_\mathrm{m}$
&
 $0.3118 \begin{array}{ll} {}_{+0.0062} & {}_{+0.013} \\ {}^{-0.0064} & {}^{-0.013} \end{array}$
&
 $0.3145 \begin{array}{ll} {}_{+0.0066} & {}_{+0.015} \\ {}^{-0.0083} & {}^{-0.015} \end{array}$
&
 $0.3135 \begin{array}{ll} {}_{+0.0066} & {}_{+0.014} \\ {}^{-0.0074} & {}^{-0.014} \end{array}$
&
 $0.3082 \begin{array}{ll} {}_{+0.0071} & {}_{+0.015} \\ {}^{-0.0073} & {}^{-0.015} \end{array}$
&
 $0.3171 \begin{array}{ll} {}_{+0.0091} & {}_{+0.019} \\ {}^{-0.0099} & {}^{-0.018} \end{array}$
\\
    \end{tabular}
  \end{footnotesize}
  \caption{
    Constraints from all cosmological models, using the MR($3$-param)
    bias model, as well as the data combinations PB (Planck + BOSS DR11)
    or PBJ (Planck + BOSS DR11 + JLA supernovae).  For each parameter,
    the mean value as well as $68\%$ and $95\%$ upper and lower bounds
    are shown.
    \label{t:constraints_1d_MR3}
  }
\end{table*}



We begin by fixing $\omega_\nu = 0.0006$, corresponding to a sum of neutrino masses of $56.6$~meV, which is the $2\sigma$ lower bound  in the normal hierarchy~\cite{Patrignani_2016}.  The first column of Table~\ref{t:constraints_1d_MR3} lists our constraints on the six vanilla parameters as well as the derived parameter $\Omegamo$.  In comparison with the Planck-only TT,TE,EE + low-$\ell$ polarization constraints of Ref.~\cite{Ade_2016m} Table~3, our measurement of
$h$ is $0.3\sigma$ higher;
$\omegac$ is $0.3\sigma$ lower;
$n_s$ is $0.4\sigma$ lower;
$\sigma_8$ is $0.8\sigma$ lower;
$\tau$ is $0.7\sigma$ lower;
and $\omegab$ is $0.2\sigma$ lower.
Our addition of BOSS DR11 data improved uncertainties in
$n_s$ by $18\%$
$h$ by $27\%$
$\omegac$ by $30\%$ and
$\omegab$ by $16\%$,
while leaving uncertainties in $\tau$ and $\sigma_8$ essentially unaffected.

Overall, our Planck + BOSS DR11 analysis is quite consistent with the Planck-only constraints of Ref.~\cite{Ade_2016m}, with all parameter shifts less than $1\sigma$, and the only shifts $\geq 0.5\sigma$ occuring in $\tau$ and $\sigma_8$ along their mutual degeneracy direction leaving $\sigma_8 \exp(-\tau)$ nearly constant.  Our constraints are also consistent with Ref.~\cite{Alam_2016}, which found $h=0.676\pm0.005$, about $0.5\sigma$ above our value, and $\Omega_\mathrm{m}=0.311\pm0.006$, in close agreement with our constraint.

\subsection{Massive neutrinos in the $\nu \Lambda$CDM model}
\label{subsec:massive_neutrinos_in_the_nuLCDM_model}

\begin{figure}[tb]
  \vskip4mm\hskip7mm~Northern
  \vskip-7mm\includegraphics[width=3.4in]{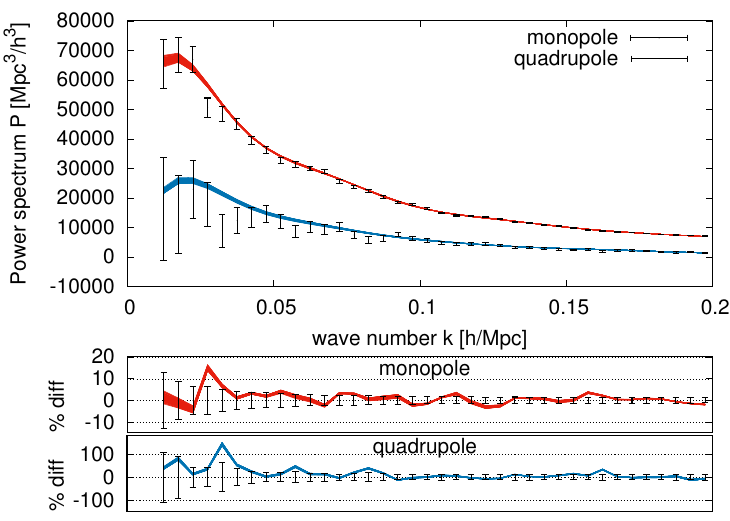}%

  \caption{
    Binned, windowed model power spectra vs.~BOSS data, for the $\nu\Lambda$CDM
    model with MR($3$-param) bias and the Planck + BOSS data combination.
    One hundred power spectra were randomly chosen from the converged portion
    of the Markov chains and plotted against the data from the northern
    sky patch.  The lower panels show the residuals.
    \label{f:bestfit_N}
  }
\end{figure}

\begin{figure}[tb]
  \vskip4mm\hskip7mm~Southern
  \vskip-7mm\includegraphics[width=3.4in]{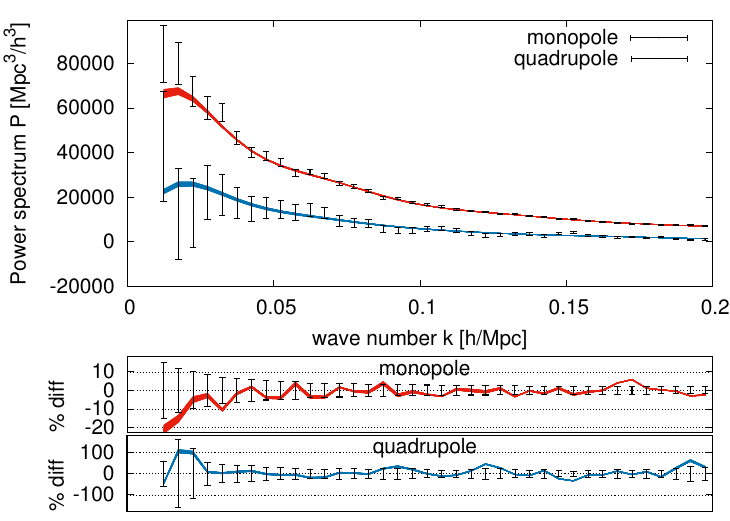}%
  \caption{
    Similar to Fig.~\ref{f:bestfit_N} but for the southern sky data.
    \label{f:bestfit_S}
    }
\end{figure}

\begin{figure}[tb]
  \includegraphics[width=3.4in]{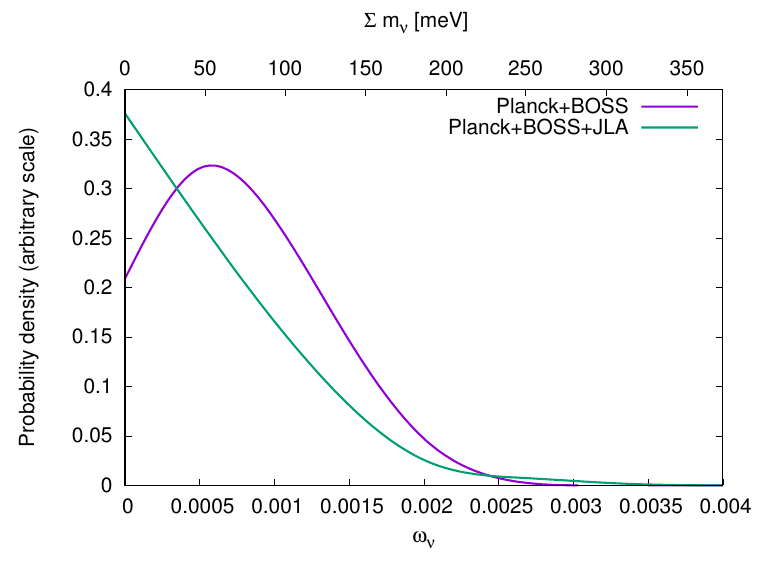}%

  \caption{
    Probability density as a function of neutrino fraction $\omeganu$ for
    $\nu\Lambda$CDM models with MR($3$-param) bias and either the
    Planck + BOSS or Planck + BOSS + JLA data combinations.  
    \label{f:nuLCDM_1d_PBJ}
  }
\end{figure}

We begin with the $\nu\Lambda$CDM model and the MR($3$-param) bias applied to the Planck + BOSS DR11 data, shown in the second column of Table~\ref{t:constraints_1d_MR3}.  Figures~\ref{f:bestfit_N} and~\ref{f:bestfit_S} plot randomly-chosen power spectra from the converged portions of our Markov chains against the BOSS data, with $\vec b$ and $\sigma_v$ set to the $\chi^2$-minimizing value for each spectrum.  For the BOSS data alone, the best-fitting power spectrum has $\chi^2 / \mathrm{d.o.f.} = 158/141 = 1.1$.

Comparing $\nu\Lambda$CDM to $\Lambda$CDM, we see that allowing $\omeganu$ to vary shifts $n_s$, $h$, $\tau$, and $\Omegamo$ toward the model preferred by the Planck-only analysis of Ref.~\cite{Ade_2016m}.  $\sigma_8$ shifts slightly lower, while $\omegac$ and $\omegab$ are only weakly affected.  A nonzero neutrino fraction $\omeganu$ is slightly preferred at the $1\sigma$ level.  Figure~\ref{f:nuLCDM_1d_PBJ} shows the marginalized probability density.

\begin{figure*}[tb]
  \includegraphics[width=2.3in]{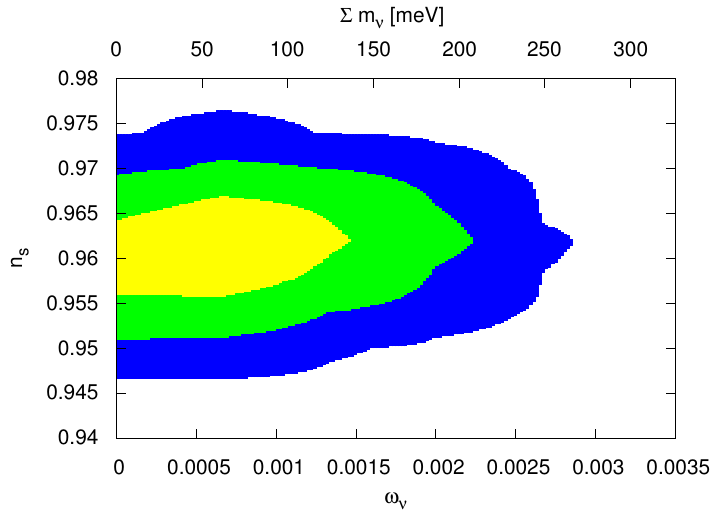}
  \includegraphics[width=2.3in]{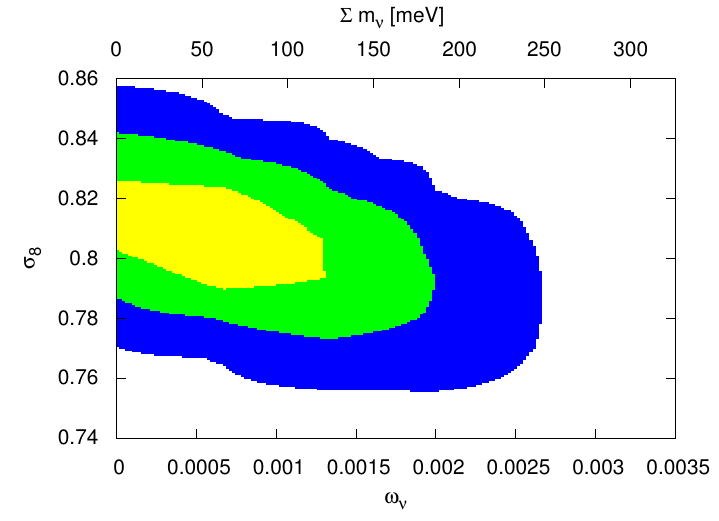}
  \includegraphics[width=2.3in]{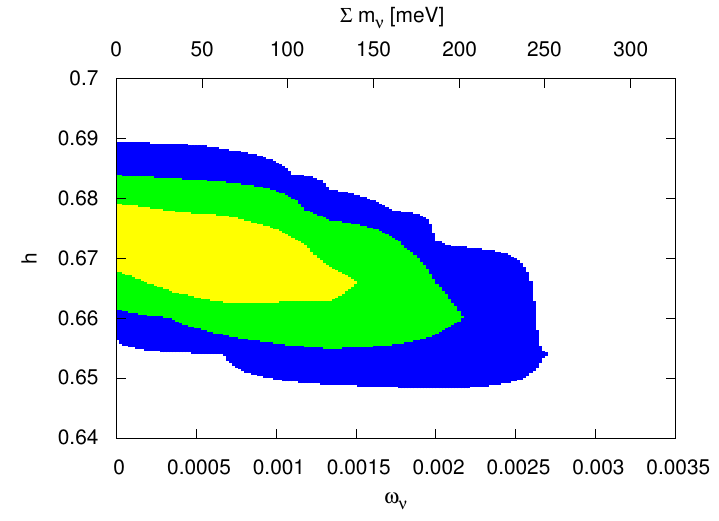}%
  
  \includegraphics[width=2.3in]{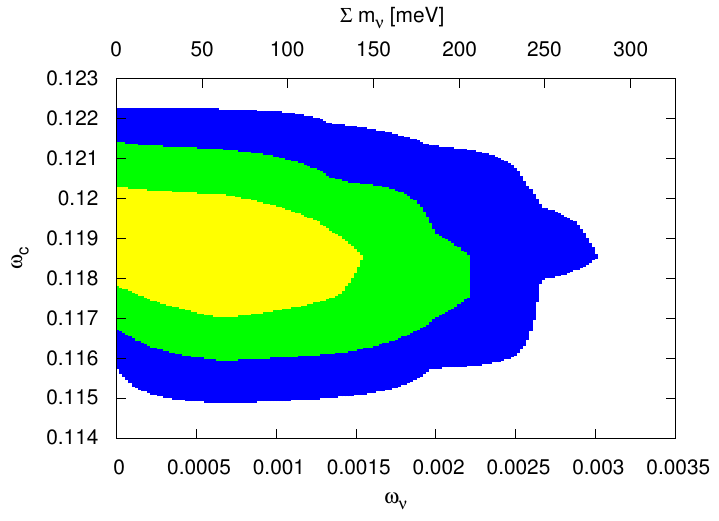}
  \includegraphics[width=2.3in]{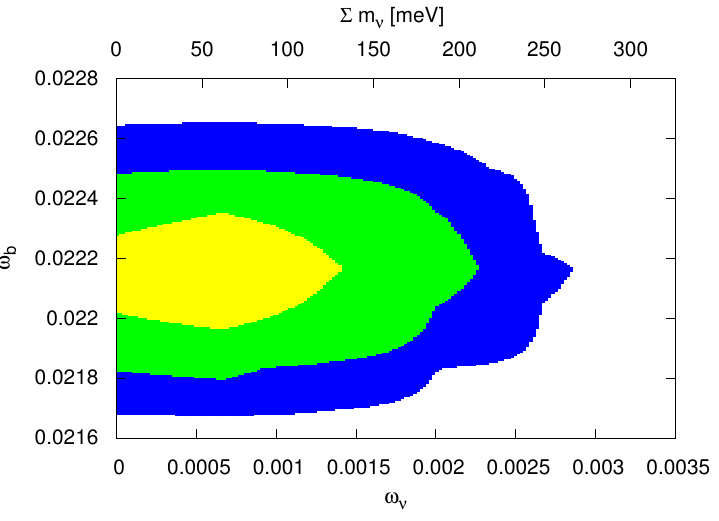}
  \includegraphics[width=2.3in]{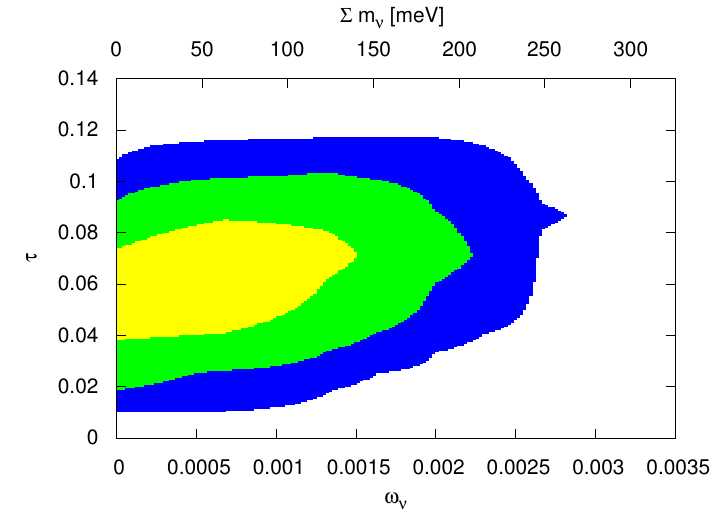}%
  
  \caption{
    Two-dimensional constraint plots for the $\nu\Lambda$CDM model with
    MR($3$-param) bias applied to Planck + BOSS DR11 data.  Light (yellow),
    medium (green), and dark (blue) regions show $68\%$, $95\%$, and $99.7\%$
    confidence regions.
    \label{f:nuLCDM_MR3_PB}
  }
\end{figure*}

Two-dimensional constraint plots for $\nu\Lambda$CDM are shown in Fig.~\ref{f:nuLCDM_MR3_PB}.  An increase in $\omeganu$ is associated with increases in $n_s$ and $\tau$ but a decrease in $h$.  Neutrinos with masses $m_\nu \lesssim 200$~meV are relativistic at decoupling, so the dominant effect of such masses is the scale-dependent suppression of power at late times.  Increasing $n_s$ to compensate for this suppression will lead to a smaller red tilt in the CMB power spectrum, explaining the preference for greater $\tau$ at greater $\omeganu$.

Although the Fourier-space power spectrum multipoles analyzed here were measured in Ref.~\cite{Beutler_2014} and applied to $\sum m_\nu$ in Ref.~\cite{Beutler_2014b}, our results disagree somewhat with those references.  For the Planck + BOSS DR11 data combination, their $95\%$ confidence upper bound $\sum m_\nu < 400$~meV is about twice our value, primarily because they find a $\approx 1.5\sigma$ preference for a higher mass $\sum m_\nu = 200$~meV.  The notable differences between the BOSS analysis of Refs.~\cite{Beutler_2014,Beutler_2014b} and the work presented here are the following.
\begin{enumerate}
\item Ref.~\cite{Beutler_2014b} treats the non-linear clustering of neutrinos
  identically to that of CDM and baryons.  This overestimates the non-linear
  neutrino power spectrum at small scales, hence underestimates the
  scale-dependent suppression of power due to neutrinos, possibly leading the
  data analysis to compensate by increasing $\sum m_\nu$.  By contrast, our
  Time-RG-based analysis treats massive neutrinos linearly as in
  Ref.~\cite{Lesgourgues_etal_2009}, while capturing the scale-dependent
  suppression of these linear neutrinos on the CDM+baryon fluid.
  \label{enum:nu_clustering}
\item The BOSS analysis constructed the galaxy power spectrum at each MCMC
  step using a template based upon phenomenological parameters such as
  $f \sigma_8$ and then used constraints on these phenomenological parameters
  to derive constraints on the cosmological parameters.  We instead took
  a brute-force approach, using Time-RG to compute the redshift-space power
  spectrum directly as a function of the cosmological parameters at each
  MCMC point and then directly computing the likelihood as in
  Appendix~\ref{sec:boss_likelihood} to constrain the cosmological parameters.
  \label{enum:pheno_params}
\item This work used the 2015 Planck data set rather than the 2013 data.
  Although $\sum m_\nu$ did not change substantially between the two Planck
  releases, the 2015 data may prefer different masses when combined with
  galaxy survey data.
\item This work assumed open priors on the bias parameters, which we consider
  appropriate for an analysis constraining new physics such as dark energy
  and non-minimal neutrino masses.  We include this as a potential difference
  since bias priors are not listed in Ref.~\cite{Beutler_2014b}.
\end{enumerate}
Reference~\cite{Beutler_2014b} studied items~\ref{enum:nu_clustering} and \ref{enum:pheno_params} in combination, finding that they caused a combined shift of $\approx 0.4\sigma$ in $f\sigma_8$.  Since a thorough study of these differences and their impact on neutrino mass constraints is well beyond the scope of this article, we simply list them here along with our results.

Figure~\ref{f:nuLCDM_1d_PBJ} and Table~\ref{t:constraints_1d_MR3} compare the marginalized neutrino mass constraints in $\nu\Lambda$CDM with and without JLA supernova data.  Adding JLA data shifts the mean $\omeganu$ somewhat lower, but the $95\%$ confidence level upper bound on $\sum m_\nu$ is virtually unchanged, from $183$~meV without JLA to $179$~meV with it.  The consistency of these CMB, supernova, and galaxy survey data sets in the $\nu\Lambda$CDM model is encouraging.

\begin{figure}[tb]
  \includegraphics[width=3.4in]{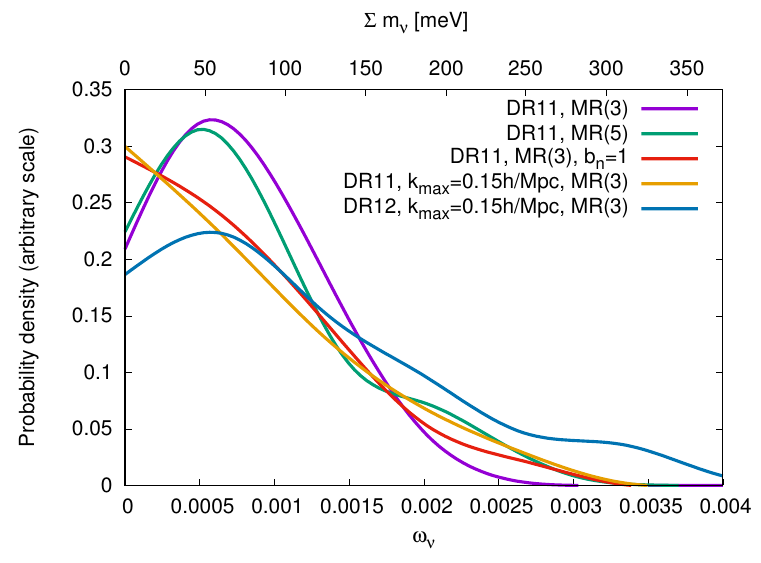}%

  \caption{
    Probability density as a function of neutrino fraction $\omeganu$ for
    $\nu\Lambda$CDM models applied to the Planck + BOSS data combination,
    using variations on the galaxy data sets and the 
    McDonald-Roy bias parameterization of Ref.~\cite{McDonald_Roy_2009}.
    \label{f:nuLCDM_1d_bias}
  }
\end{figure}


\begin{table*}[tb]
  \tabcolsep=0.02cm
  \begin{footnotesize}
    \begin{tabular}{r||c|c|c|c|c}
      &
      $\nu\Lambda$CDM
      &
      $\nu\Lambda$CDM, MR(5)
      &
      $\nu\Lambda$CDM, $b_n=1$
      &
      $\nu\Lambda$CDM, $k_\mathrm{max}\!=\!0.15\tfrac{h}{\mathrm{Mpc}}$   
      &
      $\nu\Lambda$CDM, BOSS DR12
      \\
      
      \hline
\hline
$n_s$
&
 $0.963 \begin{array}{ll} {}_{+0.0041} & {}_{+0.0076} \\ {}^{-0.0042} & {}^{-0.0082} \end{array}$
&
 $0.9633 \begin{array}{ll} {}_{+0.004} & {}_{+0.0083} \\ {}^{-0.0041} & {}^{-0.008} \end{array}$
&
 $0.9628 \begin{array}{ll} {}_{+0.0042} & {}_{+0.0083} \\ {}^{-0.004} & {}^{-0.0082} \end{array}$
&
 $0.9624 \begin{array}{ll} {}_{+0.0041} & {}_{+0.0085} \\ {}^{-0.0043} & {}^{-0.0084} \end{array}$
&
 $0.9647 \begin{array}{ll} {}_{+0.004} & {}_{+0.0086} \\ {}^{-0.0044} & {}^{-0.0086} \end{array}$
\\
\hline
$\sigma_8$
&
 $0.812 \begin{array}{ll} {}_{+0.014} & {}_{+0.028} \\ {}^{-0.014} & {}^{-0.028} \end{array}$
&
 $0.812 \begin{array}{ll} {}_{+0.018} & {}_{+0.032} \\ {}^{-0.014} & {}^{-0.034} \end{array}$
&
 $0.812 \begin{array}{ll} {}_{+0.018} & {}_{+0.031} \\ {}^{-0.014} & {}^{-0.033} \end{array}$
&
 $0.817 \begin{array}{ll} {}_{+0.019} & {}_{+0.033} \\ {}^{-0.015} & {}^{-0.037} \end{array}$
&
 $0.805 \begin{array}{ll} {}_{+0.024} & {}_{+0.036} \\ {}^{-0.014} & {}^{-0.043} \end{array}$
\\
\hline
$h$
&
 $0.6722 \begin{array}{ll} {}_{+0.0063} & {}_{+0.011} \\ {}^{-0.0053} & {}^{-0.012} \end{array}$
&
 $0.6724 \begin{array}{ll} {}_{+0.0064} & {}_{+0.012} \\ {}^{-0.005} & {}^{-0.012} \end{array}$
&
 $0.6718 \begin{array}{ll} {}_{+0.0067} & {}_{+0.012} \\ {}^{-0.0053} & {}^{-0.012} \end{array}$
&
 $0.671 \begin{array}{ll} {}_{+0.0082} & {}_{+0.014} \\ {}^{-0.0063} & {}^{-0.016} \end{array}$
&
 $0.6734 \begin{array}{ll} {}_{+0.0082} & {}_{+0.013} \\ {}^{-0.0059} & {}^{-0.014} \end{array}$
\\
\hline
$\omega_\mathrm{c}$
&
 $0.1191 \begin{array}{ll} {}_{+0.0011} & {}_{+0.0021} \\ {}^{-0.0011} & {}^{-0.0021} \end{array}$
&
 $0.1189 \begin{array}{ll} {}_{+0.0011} & {}_{+0.0023} \\ {}^{-0.0012} & {}^{-0.0023} \end{array}$
&
 $0.1192 \begin{array}{ll} {}_{+0.0012} & {}_{+0.0024} \\ {}^{-0.0012} & {}^{-0.0023} \end{array}$
&
 $0.1193 \begin{array}{ll} {}_{+0.0012} & {}_{+0.0025} \\ {}^{-0.0012} & {}^{-0.0024} \end{array}$
&
 $0.1180 \begin{array}{ll} {}_{+0.0012} & {}_{+0.0023} \\ {}^{-0.0011} & {}^{-0.0024} \end{array}$
\\
\hline
$\omega_\mathrm{b}$
&
 $0.02223 \begin{array}{ll} {}_{+0.00013} & {}_{+0.00027} \\ {}^{-0.00014} & {}^{-0.00026} \end{array}$
&
 $0.02225 \begin{array}{ll} {}_{+0.00014} & {}_{+0.00027} \\ {}^{-0.00014} & {}^{-0.00027} \end{array}$
&
 $0.02223 \begin{array}{ll} {}_{+0.00014} & {}_{+0.00027} \\ {}^{-0.00014} & {}^{-0.00027} \end{array}$
&
 $0.02223 \begin{array}{ll} {}_{+0.00014} & {}_{+0.00029} \\ {}^{-0.00014} & {}^{-0.0003} \end{array}$
&
 $0.02231 \begin{array}{ll} {}_{+0.00014} & {}_{+0.00028} \\ {}^{-0.00014} & {}^{-0.00028} \end{array}$
\\
\hline
$\omega_\nu$
&
 $0.00098 \begin{array}{ll} {}_{+0.00044} & {}_{+0.00099} \\ {}^{-0.00071} & {}^{-0.00098} \end{array}$
&
 $0.0010 \begin{array}{ll} {}_{+0.00031} & {}_{+0.0014} \\ {}^{-0.00089} & {}^{-0.001} \end{array}$
&
 $0.00098 \begin{array}{ll} {}_{+0.00024} & {}_{+0.0014} \\ {}^{-0.00098} & {}^{-0.00098} \end{array}$
&
 $0.0010 \begin{array}{ll} {}_{+0.00021} & {}_{+0.0015} \\ {}^{-0.001} & {}^{-0.001} \end{array}$
&
 $0.0014 \begin{array}{ll} {}_{+0.0004} & {}_{+0.0025} \\ {}^{-0.0013} & {}^{-0.0014} \end{array}$
\\
&
 ($<0.00197$ to $95\%$ CL)
&
 ($<0.0024$ to $95\%$ CL)
&
 ($<0.0024$ to $95\%$ CL)
&
 ($<0.0025$ to $95\%$ CL)
&
 ($<0.0039$ to $95\%$ CL)
\\
\hline
$\tau$
&
 $0.0702 \begin{array}{ll} {}_{+0.017} & {}_{+0.034} \\ {}^{-0.017} & {}^{-0.033} \end{array}$
&
 $0.0728 \begin{array}{ll} {}_{+0.017} & {}_{+0.035} \\ {}^{-0.018} & {}^{-0.035} \end{array}$
&
 $0.0691 \begin{array}{ll} {}_{+0.018} & {}_{+0.035} \\ {}^{-0.018} & {}^{-0.035} \end{array}$
&
 $0.0762 \begin{array}{ll} {}_{+0.018} & {}_{+0.036} \\ {}^{-0.018} & {}^{-0.035} \end{array}$
&
 $0.0784 \begin{array}{ll} {}_{+0.019} & {}_{+0.035} \\ {}^{-0.018} & {}^{-0.036} \end{array}$
\\
\hline
$\Omega_\mathrm{m}$
&
 $0.3145 \begin{array}{ll} {}_{+0.0066} & {}_{+0.015} \\ {}^{-0.0083} & {}^{-0.015} \end{array}$
&
 $0.3142 \begin{array}{ll} {}_{+0.0064} & {}_{+0.016} \\ {}^{-0.0084} & {}^{-0.015} \end{array}$
&
 $0.315 \begin{array}{ll} {}_{+0.007} & {}_{+0.016} \\ {}^{-0.0084} & {}^{-0.015} \end{array}$
&
 $0.3162 \begin{array}{ll} {}_{+0.0079} & {}_{+0.02} \\ {}^{-0.01} & {}^{-0.018} \end{array}$
&
 $0.3122 \begin{array}{ll} {}_{+0.0069} & {}_{+0.018} \\ {}^{-0.0099} & {}^{-0.016} \end{array}$
\\
    \end{tabular}
  \end{footnotesize}
  \caption{
    Constraints on $\nu\Lambda$CDM using variations of the McDonald-Roy
    bias model, as well as the combined Planck and BOSS data.  
    BOSS DR11 data up to $k_\mathrm{max} = 0.20~h/$Mpc and the 
    MR(3) bias model are used unless otherwise labeled.
    The fourth column uses BOSS DR11 data up to maximum wave number
    $k_\mathrm{max} = 0.15~h/$Mpc, and the fifth uses BOSS DR12 data,
    which have that same $k_\mathrm{max}$.
    For each parameter, the mean value as well as $68\%$ and $95\%$ upper 
    and lower bounds are shown.
    \label{t:constraints_1d_nuLCDM}
  }
\end{table*}



A major goal of this article is to study the robustness of the neutrino mass constraint with respect to variations on the galaxy survey analysis.  Table~\ref{t:constraints_1d_nuLCDM} and Figure~\ref{f:nuLCDM_1d_bias} compare several different galaxy data sets and bias treatments for the $\nu\Lambda$CDM model applied to the Planck + BOSS data combination.  First, consider varying the bias model, as in columns $2$ and $3$ of Table~\ref{t:constraints_1d_nuLCDM}, compared with the first column.  Allowing two extra bias parameters to vary through the MR($5$) parameterization, or fixing the neutrino bias $\bnu$ to unity rather than zero,  both weaken the $95\%$~confidence level upper bound on $\omeganu$ by $\approx 25\%$ and the $95\%$ confidence interval on $\sigma_8$ by somewhat less.  Mean parameter values are essentially unchanged, showing their robustness with respect to variations of the bias model.  The slight weakening of constraints in the MR($5$) bias model is expected, since two new parameters must be marginalized.  The similar weakening with $\bnu$ fixed to $1$ is somewhat surprising, since no new variable parameters have been added to the MR($3$) bias model.  The mean parameter values in Table~\ref{t:constraints_1d_nuLCDM} for $\bnu$ of zero and one are nearly identical, as are the maximum-likelihood points in the two sets of chains, which both have: $\ns=0.9626$; $\sigma_8=0.822$ for $\bnu=0$ vs.~$0.823$ for $\bnu=1$; $h=0.6760$ vs.~$0.6756$; $\omegac=0.1193$ vs.~$0.1194$; $\omegab=0.02223$ vs.~$0.02224$; $\omeganu=0.00055$ vs.~$0.00056$; and $\tau=0.069$ vs.~$0.070$.  Unfortunately, the maximum log-likelihoods for these points differ by only $0.005$, suggesting that the current data are not powerful enough to constrain $\bnu$ as a free parameter.  As the data improve, the neutrino contribution to the scale-dependent galaxy bias should be revisited.

Next we consider the dependence of our constraints on high-$k$ data.  Reference~\cite{Beutler_2014} recommended that their results be used up to $k_\mathrm{max}=0.20~h/$Mpc, as we have done, but also reported results for $k_\mathrm{max}=0.15~h/$Mpc.  In the fourth column of Table~\ref{t:constraints_1d_nuLCDM} we analyze BOSS DR11 data with $k_\mathrm{max}=0.15~h/$Mpc.  Not surprisingly, these constraints are weaker.  The $95\%$ confidence intervals for $\sigma_8$, $h$, and $\omeganu$ grow by $25\%-30\%$.  Although the  mean values of $\sigma_8$ and $\tau$ shift, $\Delta \ln(\sigma_8) \approx \Delta \tau$, keeping $\sigma_8 \exp(-\tau)$ approximately constant.  Aside from the weakening of constraints and this shift of $\sigma_8$ and $\tau$ along their degeneracy direction, ignoring the high-$k$ data does not qualitatively alter our results, showing that the high-$k$ and low-$k$ power spectrum data are consistent at the level of current constraints.

In the fifth column of  Table~\ref{t:constraints_1d_nuLCDM}, we consider replacing BOSS DR11 data by the more recent DR12 data of Ref.~\cite{Beutler_2017}.  We included in our DR12 analysis the hexadecapole power spectrum measured by that reference along with the monopole and quadrupole.  Aside from reducing $k_\mathrm{max}$ to $0.15~h/$Mpc, the DR12 data have a more complicated bias structure.  There are three separate redshift bins, each requiring its own independent set of bias parameters.  Additionally, the northern and southern galactic populations are different, again necessitating independent bias parameters.  Thus the MR($3$) bias model has $18$ bias parameters, three for each redshift bin and sky patch.  Combined with the open bias priors used in this analysis, DR12 has substantially worse constraints.  The $95\%$ confidence upper bound on $\sum m_\nu$ doubles from $183$~meV to $362$~meV relative to DR11 constraints with $k_\mathrm{max}=0.20~h/$Mpc, and increases by $50\%$ relative to DR11 constraints with $k_\mathrm{max}=0.15~h/$Mpc.  Bounds on the other cosmological parameters do not degrade significantly in DR12, showing that the neutrino mass constraint is especially sensitive to scale-dependent bias.


\begin{table*}[tb]
  \tabcolsep=0.02cm
  \begin{footnotesize}
    \begin{tabular}{r||c|c|c|c|c}
      &
     DR11 $k_\mathrm{max}\!=\!0.20\tfrac{h}{\mathrm{Mpc}}$
      &
     DR11 $k_\mathrm{max}\!=\!0.15\tfrac{h}{\mathrm{Mpc}}$
      &
     DR12\cite{Gil-Marin_2015} $k_\mathrm{max}\!=\!0.24\tfrac{h}{\mathrm{Mpc}}$
      &
     DR12\cite{Gil-Marin_2015} $k_\mathrm{max}\!=\!0.20\tfrac{h}{\mathrm{Mpc}}$
      &
     DR12\cite{Gil-Marin_2015} $k_\mathrm{max}\!=\!0.15\tfrac{h}{\mathrm{Mpc}}$
      \\
      
      \hline
\hline
$n_s$
&
 $0.963 \begin{array}{ll} {}_{+0.0041} & {}_{+0.0076} \\ {}^{-0.0042} & {}^{-0.0082} \end{array}$
&
 $0.9624 \begin{array}{ll} {}_{+0.0041} & {}_{+0.0085} \\ {}^{-0.0043} & {}^{-0.0084} \end{array}$
&
 $0.959 \begin{array}{ll} {}_{+0.0041} & {}_{+0.0082} \\ {}^{-0.0039} & {}^{-0.0082} \end{array}$
&
 $0.9608 \begin{array}{ll} {}_{+0.0039} & {}_{+0.0079} \\ {}^{-0.0038} & {}^{-0.0081} \end{array}$
&
 $0.9603 \begin{array}{ll} {}_{+0.0045} & {}_{+0.0084} \\ {}^{-0.0038} & {}^{-0.0084} \end{array}$
\\
\hline
$\sigma_8$
&
 $0.812 \begin{array}{ll} {}_{+0.014} & {}_{+0.028} \\ {}^{-0.014} & {}^{-0.028} \end{array}$
&
 $0.817 \begin{array}{ll} {}_{+0.019} & {}_{+0.033} \\ {}^{-0.015} & {}^{-0.037} \end{array}$
&
 $0.808 \begin{array}{ll} {}_{+0.008} & {}_{+0.025} \\ {}^{-0.017} & {}^{-0.022} \end{array}$
&
 $0.818 \begin{array}{ll} {}_{+0.013} & {}_{+0.026} \\ {}^{-0.013} & {}^{-0.026} \end{array}$
&
 $0.818 \begin{array}{ll} {}_{+0.018} & {}_{+0.03} \\ {}^{-0.012} & {}^{-0.032} \end{array}$
\\
\hline
$h$
&
 $0.6722 \begin{array}{ll} {}_{+0.0063} & {}_{+0.011} \\ {}^{-0.0053} & {}^{-0.012} \end{array}$
&
 $0.671 \begin{array}{ll} {}_{+0.0082} & {}_{+0.014} \\ {}^{-0.0063} & {}^{-0.016} \end{array}$
&
 $0.6677 \begin{array}{ll} {}_{+0.005} & {}_{+0.011} \\ {}^{-0.0059} & {}^{-0.01} \end{array}$
&
 $0.6692 \begin{array}{ll} {}_{+0.005} & {}_{+0.0094} \\ {}^{-0.0051} & {}^{-0.0092} \end{array}$
&
 $0.666 \begin{array}{ll} {}_{+0.0062} & {}_{+0.012} \\ {}^{-0.0052} & {}^{-0.012} \end{array}$
\\
\hline
$\omega_\mathrm{c}$
&
 $0.1191 \begin{array}{ll} {}_{+0.0011} & {}_{+0.0021} \\ {}^{-0.0011} & {}^{-0.0021} \end{array}$
&
 $0.1193 \begin{array}{ll} {}_{+0.0012} & {}_{+0.0025} \\ {}^{-0.0012} & {}^{-0.0024} \end{array}$
&
 $0.1208 \begin{array}{ll} {}_{+0.0012} & {}_{+0.0024} \\ {}^{-0.0013} & {}^{-0.0023} \end{array}$
&
 $0.1204 \begin{array}{ll} {}_{+0.0011} & {}_{+0.0022} \\ {}^{-0.0012} & {}^{-0.0022} \end{array}$
&
 $0.1206 \begin{array}{ll} {}_{+0.0011} & {}_{+0.0025} \\ {}^{-0.0013} & {}^{-0.0025} \end{array}$
\\
\hline
$\omega_\mathrm{b}$
&
 $0.02223 \begin{array}{ll} {}_{+0.00013} & {}_{+0.00027} \\ {}^{-0.00014} & {}^{-0.00026} \end{array}$
&
 $0.02223 \begin{array}{ll} {}_{+0.00014} & {}_{+0.00029} \\ {}^{-0.00014} & {}^{-0.0003} \end{array}$
&
 $0.02209 \begin{array}{ll} {}_{+0.00015} & {}_{+0.00027} \\ {}^{-0.00014} & {}^{-0.00029} \end{array}$
&
 $0.02213 \begin{array}{ll} {}_{+0.00013} & {}_{+0.00027} \\ {}^{-0.00013} & {}^{-0.00027} \end{array}$
&
 $0.02212 \begin{array}{ll} {}_{+0.00015} & {}_{+0.00028} \\ {}^{-0.00013} & {}^{-0.00028} \end{array}$
\\
\hline
$\omega_\nu$
&
 $0.00098 \begin{array}{ll} {}_{+0.00044} & {}_{+0.00099} \\ {}^{-0.00071} & {}^{-0.00098} \end{array}$
&
 $0.001 \begin{array}{ll} {}_{+0.00021} & {}_{+0.0015} \\ {}^{-0.001} & {}^{-0.001} \end{array}$
&
 $0.00058 \begin{array}{ll} {}_{+0.00013} & {}_{+0.00082} \\ {}^{-0.00058} & {}^{-0.00058} \end{array}$
&
 $0.00064 \begin{array}{ll} {}_{+0.00013} & {}_{+0.00086} \\ {}^{-0.00064} & {}^{-0.00064} \end{array}$
&
 $0.00092 \begin{array}{ll} {}_{+0.00058} & {}_{+0.0011} \\ {}^{-0.00092} & {}^{-0.00092} \end{array}$
\\
&
 ($<0.00197$ to $95\%$ CL)
&
 ($<0.0025$ to $95\%$ CL)
&
 ($<0.0014$ to $95\%$ CL)
&
 ($<0.0015$ to $95\%$ CL)
&
 ($<0.0021$ to $95\%$ CL)
\\
\hline
$\tau$
&
 $0.0702 \begin{array}{ll} {}_{+0.017} & {}_{+0.034} \\ {}^{-0.017} & {}^{-0.033} \end{array}$
&
 $0.0762 \begin{array}{ll} {}_{+0.018} & {}_{+0.036} \\ {}^{-0.018} & {}^{-0.035} \end{array}$
&
 $0.0463 \begin{array}{ll} {}_{+0.016} & {}_{+0.033} \\ {}^{-0.02} & {}^{-0.031} \end{array}$
&
 $0.062 \begin{array}{ll} {}_{+0.014} & {}_{+0.031} \\ {}^{-0.017} & {}^{-0.031} \end{array}$
&
 $0.0682 \begin{array}{ll} {}_{+0.015} & {}_{+0.034} \\ {}^{-0.018} & {}^{-0.032} \end{array}$
\\
\hline
$\Omega_\mathrm{m}$
&
 $0.3145 \begin{array}{ll} {}_{+0.0066} & {}_{+0.015} \\ {}^{-0.0083} & {}^{-0.015} \end{array}$
&
 $0.3162 \begin{array}{ll} {}_{+0.0079} & {}_{+0.02} \\ {}^{-0.01} & {}^{-0.018} \end{array}$
&
 $0.3213 \begin{array}{ll} {}_{+0.0079} & {}_{+0.014} \\ {}^{-0.0078} & {}^{-0.015} \end{array}$
&
 $0.3192 \begin{array}{ll} {}_{+0.0066} & {}_{+0.013} \\ {}^{-0.007} & {}^{-0.013} \end{array}$
&
 $0.3235 \begin{array}{ll} {}_{+0.0071} & {}_{+0.016} \\ {}^{-0.0087} & {}^{-0.016} \end{array}$
\\
    \end{tabular}
  \end{footnotesize}
  \caption{
    Constraints on $\nu\Lambda$CDM using the MR(3) bias model, the 
    Planck data, and two different BOSS data sets.  BOSS DR11 data 
    are compared with the BOSS DR12 data of Ref.~\cite{Gil-Marin_2015}, 
    which kept the CMASS and LOWZ galaxy populations separate, 
    simplifying the galaxy bias analysis. The first two  columns analyze 
    DR11 data up to maximum wave numbers of $k_\mathrm{max} = 0.20~h/$Mpc 
    and $0.15~h/$Mpc, respectively.  The third, fourth, and fifth columns 
    analyze the DR12 data up to $k_\mathrm{max} = 0.24~h/$Mpc, $0.20~h/$Mpc, 
    and $0.15~h/$Mpc, respectively.  
    For each parameter, the mean value as well as $68\%$ and $95\%$ upper 
    and lower bounds are shown.
    \label{t:constraints_1d_nuLCDM_DR12GM}
  }
\end{table*}



In Table~\ref{t:constraints_1d_nuLCDM_DR12GM}, we study the earlier BOSS DR12 data of Ref.~\cite{Gil-Marin_2015}. Since that reference kept the LOWZ and CMASS galaxy populations separate, their analysis requires fewer bias parameters than the DR12 data of Ref.~\cite{Beutler_2017}, and the resulting neutrino mass constraints are tighter.  However, our constraints using Ref.~\cite{Gil-Marin_2015} data exhibit a sensitivity to high wave numbers not seen in the DR11 data.  As the maximum wave number analyzed is reduced from  $k_\mathrm{max}=0.24~h/$Mpc in the third column of Table~\ref{t:constraints_1d_nuLCDM_DR12GM} to $0.20~h/$Mpc in the fourth column and $0.15~h/$Mpc in the fifth column, the mean values of $\sigma_8$ and $\tau$ drift by $\approx 1\sigma$, and the mean and upper bound on $\omega_\nu$ increase by $\approx 50\%$.  These drifts are qualitatively consistent with Fig.~9 of Ref.~\cite{Gil-Marin_2015}, which finds $0.5\sigma-1\sigma$ drifts in $f\sigma_8$ and the Alcock-Paczynski parameter $\alpha_\parallel$ as $k_\mathrm{max}$ is varied.  Since our earlier discussion shows that neutrino mass constraints are especially sensitive to scale-dependent systematics, we choose to use the BOSS DR11 data henceforth.

Lastly, we compare our neutrino mass constraints to the literature.  The BOSS DR12 constraint of Ref.~\cite{Alam_2016}, which used Ref.~\cite{Beutler_2017}, found $\sum m_\nu < 160$~meV at the 95\% confidence level, corresponding to $\omega_\nu < 0.0017$, comparable to our constraint.  The earlier BOSS DR7 analysis of Ref.~\cite{Cuesta_2015}, which supplemented halo power spectrum constraints with measurements of the BAO distance scale at a range of redshifts, found $\sum m_\nu < 130$~meV, or $\omega_\nu < 0.0014$.  Their constraint on the remaining parameters are consistent with ours at the $\approx 1\sigma$ level despite using substantially different data.

\subsection{Dark energy and its evolution}
\label{subsec:dark_energy_and_its_evolution}

Finally, we consider the effect of dark energy with an evolving equation of state on the massive neutrino constraint, and vice versa.  Since Planck and BOSS alone are insufficient for constraining the evolution of the dark energy equation of state, we exclusively consider the Planck + BOSS + JLA data combination here.  We parameterize the dark energy as a non-clustering perfect fluid with unit sound speed and equation of state $w(z) = w_0 + w_a z / (1+z)$, following Refs.~\cite{Chevalier_Polarski_2001,Linder_2003}.

\begin{figure}[tb]
  \includegraphics[width=3.4in]{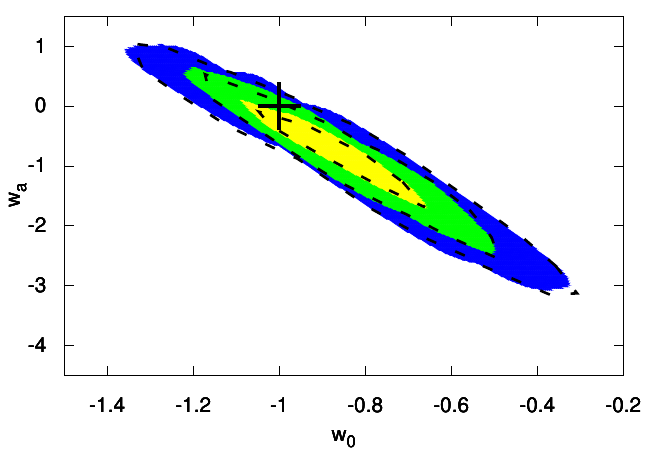}%
  
  \includegraphics[width=3.4in]{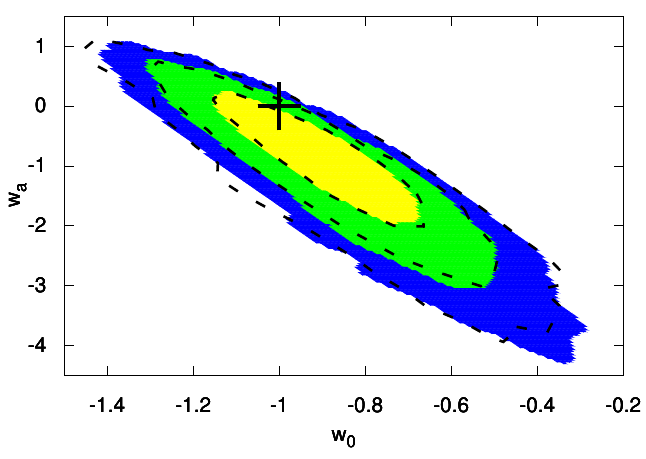}%
  \caption{
    Constraints on the dark energy equation of state $w(z) = w_0 + w_a z /(1+z)$
    in the $w$CDM model (top) and the $\nu w$CDM model (bottom). Light (yellow),
    medium (green), and dark (blue) regions show $68\%$, $95\%$, and $99.7\%$
    confidence regions. The ``+''
    symbol identifies the cosmological constant, $w(z) = -1$, which lies
    within the $95\%$ confidence region in both cases.
    Dashed lines show the same contours using the MR($5$) bias model.
    \label{f:w0_wa}
  }
\end{figure}

Figure~\ref{f:w0_wa} shows constraints on $w_0$ and $w_a$ with $\omeganu$ fixed (top) and variable (bottom). With $\omeganu=0.0006$ fixed, our constraints are stronger than those of Ref.~\cite{Ade_2016n} using Planck + BAO/RSD + weak lensing, but weaker than those using Planck + BAO + supernovae + $H_0$ measurements.  Our constraints are also similar at the $0.5\sigma$ level to BOSS DR12 constraints of Ref.~\cite{Alam_2016}, which found $w_0=-0.91\pm0.10$ and $w_a=-0.39\pm0.34$.  Surprisingly, the widths of the $95\%$ confidence allowed ranges for $w_0$ and $w_a$ have barely changed from the author's earlier work in Ref.~\cite{Upadhye_Ishak_Steinhardt_2005}, from $0.7$ to $0.6$ for $w_0$ and from $2.6$ to $2.5$ for $w_a$.  We attribute this to the fact that the earlier data preferred parameters near the boundary $w_0+w_a=0$ and were therefore artificially truncated by prior constraints.  That reference showed that, near this boundary, constraints were dependent on the choice of $w(z)$ parameterization, and that a parameterization allowing larger low-$z$ derivatives opened up a larger range of allowed parameters.

\begin{figure}[tb]
  \includegraphics[width=3.4in]{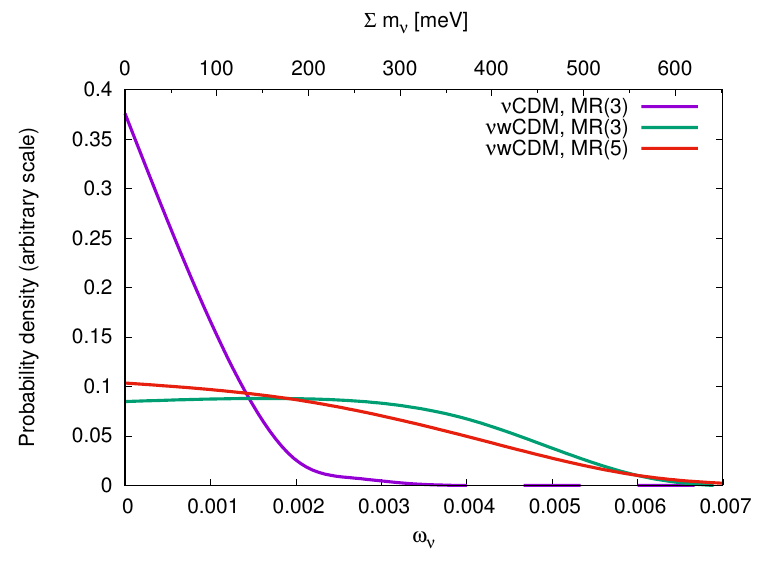}%

  \caption{
    Probability density vs.~$\omeganu$ for the $\nu\Lambda$CDM and $\nu w$CDM
    models with MR($3$-param) bias, using Planck + BOSS DR11 + JLA data.
    \label{f:omega_nu_w0wa}
  }
\end{figure}


\begin{table*}[tb]
  \tabcolsep=0.02cm
  \begin{footnotesize}
    \begin{tabular}{r||c|c|c|c|c}
      &
      $\nu\Lambda$CDM, MR(3), PBJ
      &
      $w$CDM, MR(3), PBJ
      &
      $w$CDM, MR(5), PBJ
      &
      $\nu w$CDM, MR(3), PBJ
      &
      $\nu w$CDM, MR(5), PBJ
      \\
      
      \hline
\hline
$n_s$
&
 $0.9627 \begin{array}{ll} {}_{+0.0039} & {}_{+0.008} \\ {}^{-0.004} & {}^{-0.008} \end{array}$
&
 $0.9622 \begin{array}{ll} {}_{+0.0041} & {}_{+0.0085} \\ {}^{-0.0044} & {}^{-0.0084} \end{array}$
&
 $0.9618 \begin{array}{ll} {}_{+0.0042} & {}_{+0.0089} \\ {}^{-0.0046} & {}^{-0.0086} \end{array}$
&
 $0.9613 \begin{array}{ll} {}_{+0.0044} & {}_{+0.0091} \\ {}^{-0.0045} & {}^{-0.009} \end{array}$
&
 $0.9612 \begin{array}{ll} {}_{+0.0045} & {}_{+0.0091} \\ {}^{-0.0045} & {}^{-0.009} \end{array}$
\\
\hline
$\sigma_8$
&
 $0.815 \begin{array}{ll} {}_{+0.016} & {}_{+0.029} \\ {}^{-0.014} & {}^{-0.03} \end{array}$
&
 $0.825 \begin{array}{ll} {}_{+0.015} & {}_{+0.028} \\ {}^{-0.013} & {}^{-0.029} \end{array}$
&
 $0.828 \begin{array}{ll} {}_{+0.014} & {}_{+0.03} \\ {}^{-0.015} & {}^{-0.029} \end{array}$
&
 $0.799 \begin{array}{ll} {}_{+0.024} & {}_{+0.044} \\ {}^{-0.024} & {}^{-0.044} \end{array}$
&
 $0.804 \begin{array}{ll} {}_{+0.027} & {}_{+0.042} \\ {}^{-0.02} & {}^{-0.045} \end{array}$
\\
\hline
$h$
&
 $0.6732 \begin{array}{ll} {}_{+0.0058} & {}_{+0.011} \\ {}^{-0.0049} & {}^{-0.011} \end{array}$
&
 $0.6788 \begin{array}{ll} {}_{+0.0074} & {}_{+0.015} \\ {}^{-0.0076} & {}^{-0.015} \end{array}$
&
 $0.6796 \begin{array}{ll} {}_{+0.0067} & {}_{+0.014} \\ {}^{-0.0068} & {}^{-0.014} \end{array}$
&
 $0.6753 \begin{array}{ll} {}_{+0.0073} & {}_{+0.015} \\ {}^{-0.0077} & {}^{-0.015} \end{array}$
&
 $0.6771 \begin{array}{ll} {}_{+0.0069} & {}_{+0.014} \\ {}^{-0.0066} & {}^{-0.014} \end{array}$
\\
\hline
$\omega_\mathrm{c}$
&
 $0.1193 \begin{array}{ll} {}_{+0.0012} & {}_{+0.0022} \\ {}^{-0.0011} & {}^{-0.0022} \end{array}$
&
 $0.1194 \begin{array}{ll} {}_{+0.0013} & {}_{+0.0025} \\ {}^{-0.0014} & {}^{-0.0026} \end{array}$
&
 $0.1195 \begin{array}{ll} {}_{+0.0014} & {}_{+0.0026} \\ {}^{-0.0012} & {}^{-0.0026} \end{array}$
&
 $0.1197 \begin{array}{ll} {}_{+0.0013} & {}_{+0.0027} \\ {}^{-0.0014} & {}^{-0.0026} \end{array}$
&
 $0.1198 \begin{array}{ll} {}_{+0.0014} & {}_{+0.0026} \\ {}^{-0.0013} & {}^{-0.0027} \end{array}$
\\
\hline
$\omega_\mathrm{b}$
&
 $0.02223 \begin{array}{ll} {}_{+0.00014} & {}_{+0.00027} \\ {}^{-0.00014} & {}^{-0.00027} \end{array}$
&
 $0.02222 \begin{array}{ll} {}_{+0.00014} & {}_{+0.00028} \\ {}^{-0.00014} & {}^{-0.00029} \end{array}$
&
 $0.02221 \begin{array}{ll} {}_{+0.00014} & {}_{+0.00029} \\ {}^{-0.00015} & {}^{-0.00028} \end{array}$
&
 $0.02217 \begin{array}{ll} {}_{+0.00016} & {}_{+0.00029} \\ {}^{-0.00015} & {}^{-0.00031} \end{array}$
&
 $0.02216 \begin{array}{ll} {}_{+0.00016} & {}_{+0.0003} \\ {}^{-0.00014} & {}^{-0.00029} \end{array}$
\\
\hline
$\omega_\nu$
&
 $0.00078 \begin{array}{ll} {}_{+0.00016} & {}_{+0.0011} \\ {}^{-0.00078} & {}^{-0.00078} \end{array}$
&
&
&
 $0.0029 \begin{array}{ll} {}_{+0.0018} & {}_{+0.0029} \\ {}^{-0.0029} & {}^{-0.0029} \end{array}$
&
 $0.0026 \begin{array}{ll} {}_{+0.0014} & {}_{+0.0035} \\ {}^{-0.0026} & {}^{-0.0026} \end{array}$
\\
&
 ($<0.00192$ to $95\%$ CL)
&
&
&
 ($<0.0058$ to $95\%$ CL)
&
 ($<0.0061$ to $95\%$ CL)
\\
\hline
$w_0$
&
&
 $-0.87 \begin{array}{ll} {}_{+0.15} & {}_{+0.32} \\ {}^{-0.16} & {}^{-0.29} \end{array}$
&
 $-0.86 \begin{array}{ll} {}_{+0.14} & {}_{+0.3} \\ {}^{-0.15} & {}^{-0.3} \end{array}$
&
 $-0.88 \begin{array}{ll} {}_{+0.15} & {}_{+0.32} \\ {}^{-0.17} & {}^{-0.33} \end{array}$
&
 $-0.88 \begin{array}{ll} {}_{+0.16} & {}_{+0.32} \\ {}^{-0.17} & {}^{-0.32} \end{array}$
\\
\hline
$w_a$
&
&
 $-0.61 \begin{array}{ll} {}_{+0.76} & {}_{+1.2} \\ {}^{-0.61} & {}^{-1.3} \end{array}$
&
 $-0.68 \begin{array}{ll} {}_{+0.69} & {}_{+1.2} \\ {}^{-0.52} & {}^{-1.2} \end{array}$
&
 $-0.90 \begin{array}{ll} {}_{+0.94} & {}_{+1.5} \\ {}^{-0.6} & {}^{-1.6} \end{array}$
&
 $-0.93 \begin{array}{ll} {}_{+0.89} & {}_{+1.4} \\ {}^{-0.64} & {}^{-1.6} \end{array}$
\\
\hline
$\tau$
&
 $0.0676 \begin{array}{ll} {}_{+0.016} & {}_{+0.034} \\ {}^{-0.018} & {}^{-0.033} \end{array}$
&
 $0.0648 \begin{array}{ll} {}_{+0.017} & {}_{+0.036} \\ {}^{-0.016} & {}^{-0.032} \end{array}$
&
 $0.0653 \begin{array}{ll} {}_{+0.017} & {}_{+0.033} \\ {}^{-0.017} & {}^{-0.032} \end{array}$
&
 $0.0728 \begin{array}{ll} {}_{+0.019} & {}_{+0.035} \\ {}^{-0.018} & {}^{-0.036} \end{array}$
&
 $0.0717 \begin{array}{ll} {}_{+0.018} & {}_{+0.035} \\ {}^{-0.018} & {}^{-0.035} \end{array}$
\\
\hline
$\Omega_\mathrm{m}$
&
 $0.3135 \begin{array}{ll} {}_{+0.0066} & {}_{+0.014} \\ {}^{-0.0074} & {}^{-0.014} \end{array}$
&
 $0.3082 \begin{array}{ll} {}_{+0.0071} & {}_{+0.015} \\ {}^{-0.0073} & {}^{-0.015} \end{array}$
&
 $0.3076 \begin{array}{ll} {}_{+0.0064} & {}_{+0.014} \\ {}^{-0.007} & {}^{-0.014} \end{array}$
&
 $0.3171 \begin{array}{ll} {}_{+0.0091} & {}_{+0.019} \\ {}^{-0.0099} & {}^{-0.018} \end{array}$
&
 $0.315 \begin{array}{ll} {}_{+0.008} & {}_{+0.018} \\ {}^{-0.0095} & {}^{-0.017} \end{array}$
\\
    \end{tabular}
  \end{footnotesize}
  \caption{
    Constraints on cosmological models using the Planck + BOSS DR11 
    + JLA supernovae data combination and either the MR($3$-param)
    or the the MR($5$-param) bias models.  
    For each parameter, the mean value as well as $68\%$ and $95\%$ 
    upper and lower bounds are shown.
    \label{t:constraints_1d_PBJ}
  }
\end{table*}



Consider the effect on the other cosmological parameters of allowing $\omeganu$ to vary.  For the MR($3$) bias model, compare the second and fourth columns of Table~\ref{t:constraints_1d_PBJ}.  The $95\%$ confidence bounds on $\sigma_8$ grow by $\approx 50\%$, and those on $w_a$ by $\approx 25\%$. Additionally, allowing $\omeganu$ to vary drops $\sigma_8$ by $2\sigma$, $h$ and $w_a$ by $0.5\sigma$, and raises $\tau$ by $0.5\sigma$, along with smaller shifts in other parameters.  The derived parameter $\Omegam$ also rises by $1.3\sigma$.

Next, we study the effect of dark energy on neutrino mass constraints by comparing the first and fourth columns of Table~\ref{t:constraints_1d_PBJ}.  We find a severe, factor-of-three degradation in the neutrino mass constraint when $w_0$ and $w_a$ are allowed to vary, as illustrated in Fig.~\ref{f:omega_nu_w0wa}. We are not aware of a similar test using the current data in the literature, although Ref.~\cite{Alam_2016} found a factor-of-two degradation in the $\sum m_\nu$ bound when $w_0$ and the spatial curvature were allowed to vary.  

Qualitatively, varying the neutrino mass will modify ($i$) the comoving distance $\chi(z)$ at high $z$, and hence the angular scale of the acoustic oscillations; ($ii$)  the high-$z$ growth factor $D(z)$, and hence the relationship between $\sigma_8$ and the amplitude of the initial power spectrum; and ($iii$) the small-scale suppression of the galaxy power spectrum.  However, allowing dark energy to modify $\chi(z)$ and $D(z)$ at low redshifts also weakens the link between the early- and late-time geometry and power.  Moreover, non-linear corrections to the power spectrum mean that varying $w(z)$ affects $P(k)$ differently at different scales, as can be seen in, {\em e.g.}, Ref.~\cite{Upadhye_2016}, so the substantial variations in $w$ allowed by current data may also weaken constraints from the scale-dependent growth of structure.  In principle, future tomographic surveys of large-scale structure should be able to map out $\chi(z)$ and $D(z)$ for $z \lesssim 1$, breaking the degeneracy between the $z \sim 1$ effects of dark energy and the $z \gtrsim 100$ effects of massive neutrinos.

Finally, we compare the second column to the third and the fourth to the fifth in Table~\ref{t:constraints_1d_PBJ} to explore the effects of adding new bias parameters on the $w$CDM and $\nu w$CDM constraints.  In both cases, changing the bias model from MR($3$) to MR($5$) has only a negligible effect on the mean values and $95\%$ confidence intervals of the cosmological parameters.  This insensitivity is also evident in Fig.~\ref{f:w0_wa}, which plots MR($3$) contours using shaded regions and MR($5$) contours using dashed lines.  

\section{Conclusions}
\label{sec:conclusions}

We have used the amplitude and shape of the BOSS DR11 redshift-space power spectrum, in combination with CMB and supernova data, to constrain the sum of neutrino masses and the evolution of the dark energy equation of state.  Table~\ref{t:constraints_1d_MR3} in Section~\ref{sec:results_and_discussion} lists our main constraints, including a $95\%$ confidence level upper bound $\omeganu < 0.00197$, implying $\sum m_\nu < 183$~meV.  We find that dark energy is consistent with a cosmological constant, but allows a wide range of equations of state, including rapidly-varying equations of state with derivatives $w_a \gtrsim 0.5$ or $w_a \lesssim -2.5$.  Allowing for the simultaneous variation of the neutrino mass and the equation of state weakens both sets of constraints, with the neutrino mass bound rising to $\sum m_\nu < 540$~meV at $95\%$~CL, as shown in Figs.~\ref{f:w0_wa} and \ref{f:omega_nu_w0wa} as well as Table~\ref{t:constraints_1d_PBJ}.  Thus our uncertainty in the nature of the dark energy is currently the single greatest obstacle to cosmological constraints on the sum of neutrino masses.

Additionally, we have studied a range of bias models in order to assess the dependence of neutrino mass constraints on galaxy bias.  Section~\ref{subsec:galaxies_as_biased_tracers} compares the top-down scale-dependent bias models used here to a bottom-up HOD approach based upon N-body simulations, and finds a broad agreement over the range of scales relevant to current data.  Moreover, including more bias parameters improves the fit over a larger range of scales.  In Section~\ref{sec:results_and_discussion}, Table~\ref{t:constraints_1d_nuLCDM} takes a more detailed look at the effects on $\nu\Lambda$CDM parameter constraints of the choice of bias model and galaxy survey data set.  Allowing more bias parameters to vary, or allowing the galaxy power spectrum to depend directly on the neutrino power spectrum,  weakens neutrino mass constraints without substantially changing the mean values of any cosmological parameter.  Similarly, discarding the smallest-scale BOSS data by choosing a maximum wave number $k_\mathrm{max}$ weakens the $\omeganu$ bound without significantly shifting the cosmological parameter values.  Cosmological parameters in the $w$CDM and $\nu w$CDM models are likewise robust with respect to choice of bias model, as shown in Table~\ref{t:constraints_1d_PBJ}.  Thus, at the level of the current data, today's state-of-the-art galaxy bias models are powerful enough to provide robust constraints on the ``vanilla'' set of $\Lambda$CDM parameters as well as extensions including  massive neutrinos and dark energy.

Finally, this article represents the first direct application of FAST-PT-enhanced redshift-space perturbation theory with a scale-dependent growth characteristic of massive neutrinos to the analysis of galaxy survey data.  Using convolution and fast Fourier transform methods, we sped up Time-RG perturbation theory, designed specifically for massive neutrino models, by a factor of forty, making it computationally competitive with other perturbation theories and allowing its use in a MCMC analysis.~\footnote{Code available from author on request.}  We also applied these numerical methods to the convolution integrals arising in the redshift-space power spectrum corrections of Taruya, Nishimichi, and Saito, Ref.~\cite{Taruya_etal_2010}, as well as the scale-dependent bias integrals of McDonald and Roy, Ref.~\cite{McDonald_Roy_2009}.  This work demonstrates the strength of FFT techniques for speeding up perturbation theory in real-world applications, and opens up the possibility of employing advanced perturbative methods more broadly in cosmological analyses.

\appendix
\section{Scale-dependent bias}
\label{sec:scale-dependent_bias}

In the irrotational-velocity approximation used here, the matter power spectrum depends upon correlation functions of the scalar quantities $\delta$ and $\theta$.  However, observations measure overdensities in the galaxy field, and galaxies  trace matter in a biased, scale-dependent matter.  In Reference~\cite{McDonald_Roy_2009}, McDonald and Roy use $\delta$, the velocity field $v_i$, the gravitational potential $\Phi$, and their derivatives to construct the most general set of scalar quantities up to third order in the perturbations.  Letting the observed galaxy overdensity $\deltag$ be an arbitrary linear combination of these terms, and the galaxy velocity $\thetg = b_v \theta$ be a biased tracer of the total matter velocity, the galaxy density and velocity power spectra may be written:
\begin{eqnarray}
\Pgdd\!\!\!
&=&
\bd^2 \Pdd
+ 2 \bd \bdd P_{\delta\delta^2}
+ 2 \bd \bss P_{\delta s^2}
+ \bdd^2 P_{\delta^2\delta^2}
\label{e:Pgdd}
\nonumber \\
&~&
+ 2 \bdd \bss P_{\delta^2s^2}
+ \bss^2 P_{s^2s^2}
+ 2 \bd \btnl P_{3\mathrm{nl}}
+ N
\\
\Pgdt\!\!\!
&=&
\bd \bv \Pdt
\!+\! \bdd \bv P_{\theta\delta^2}
\!+\! \bss \bv P_{\theta s^2}
\!+\! \btnl \bv f P_{3\mathrm{nl}}
\qquad
\label{e:Pgdt}
\\
\Pgtt\!\!\!
&=&
\bv^2 \Ptt
\label{e:Pgtt}
\end{eqnarray}
where
\begin{eqnarray}
P_{\delta\delta^2}\!\!
&=&\!\!\!
\int\!\!\! \frac{d^3q}{(2\pi)^3} \Plin(q) \Plin(p_-) \Fspt(\vec q,\vec p_-)
\label{e:Pdd2}
\\
P_{\theta\delta^2}\!\!
&=&\!\!\!
\int\!\!\! \frac{d^3q}{(2\pi)^3} f \Plin(q) \Plin(p_-) \Gspt(\vec q,\vec p_-)
\label{e:Ptd2}
\\
P_{\delta s^2}\!\!
&=&\!\!\!
\int\!\!\! \frac{d^3q}{(2\pi)^3} 
\Plin(q) \Plin(p_-) \Fspt(\vec q,\vec p_-) S^{(2)}(\vec q, \vec p_-)
\label{e:Pds2}
\\
P_{\theta s^2}\!\!
&=&\!\!\!
\int\!\!\! \frac{d^3q}{(2\pi)^3} 
f \Plin(q) \Plin(p_-) \Gspt(\vec q,\vec p_-) S^{(2)}(\vec q, \vec p_-)
\label{e:Pts2}
\\
P_{\delta^2\delta^2}\!\!
&=&\!\!\!
\int\!\!\! \frac{d^3q}{(2\pi)^3} 
\frac{\Plin(q)}{2}\left[\Plin(p_-)-\Plin(q)\right]
\label{e:Pd2d2}
\\
P_{\delta^2s^2}\!\!
&=&\!\!\!
\int\!\!\! \frac{d^3q}{(2\pi)^3} 
\frac{\Plin(q)}{2} \!\! \left[\Plin(p_-)S^{(2)}(\vec q,\vec p_-) 
  \!-\! \tfrac{2}{3}\Plin(q)\right]
\label{e:Pd2s2}
\\
P_{s^2s^2}\!\!
&=&\!\!\!
\int\!\!\! \frac{d^3q}{(2\pi)^3} 
\!\!
\frac{\Plin(q)}{2} \!\! \left[\! \Plin(p_-)S^{(2)}\!(\vec q,\vec p_-)^2 
  \!\!-\! \tfrac{4}{9}\Plin(q) \! \right]\qquad
\label{e:Ps2s2}
\\
\frac{P_{3\mathrm{nl}}}{\Plin}\!\!
&=&\!
\frac{105}{16}
\!\!\int\!\!\! \frac{d^3q}{(2\pi)^3} 
\Plin(q) 
\left[ K^{(2)}(\vec q,\vec k) + \tfrac{8}{63} \right]
\label{e:P3nl}
\end{eqnarray}
where all power spectra on the left hand sides are functions of the wave number $\vec k$; $\Pdd$, $\Pdt$, and $\Ptt$ are the non-linear power spectra from perturbation theory; we have defined $\vec p_- \equiv \vec k - \vec q$; and the quantities $\Fspt$, $\Gspt$, $S^{(2)}$, and $K^{(2)}$ are given by
\begin{eqnarray}
\Fspt(\vec p, \vec q)
&=&
\frac{5}{7} 
+ \frac{\vec p \cdot \vec q}{2 p q}\left(\frac{p}{q} + \frac{q}{p}\right)
+ \frac{2}{7}\left(\frac{\vec p \cdot \vec q}{pq}\right)^2
\\
\Gspt(\vec p, \vec q)
&=&
\frac{3}{7} 
+ \frac{\vec p \cdot \vec q}{2 p q}\left(\frac{p}{q} + \frac{q}{p}\right)
+ \frac{4}{7}\left(\frac{\vec p \cdot \vec q}{pq}\right)^2
\\
S^{(2)}(\vec p, \vec q)
&=&
\left(\frac{\vec p \cdot \vec q}{p q}\right)^2 - \frac{1}{3}
\\
K^{(2)}(\vec p, \vec q)
&=&
\frac{2}{7}\left[ S^{(2)}(-\vec p,\vec q) - \frac{2}{3}\right] 
S^{(2)}(\vec p, \vec q - \vec p). \qquad
\end{eqnarray}
The bias is therefore described by six parameters, $\vec b = (\bd, \bdd, \bss, \btnl, N, \bv)$, the first five of which affect the galaxy density-density power spectrum $\Pgdd$.

\begin{table*}[tb]
  \tabcolsep=0.02cm
  \begin{footnotesize}
    \begin{tabular}{r||c|c|c|c|c|c|c|c|c|c|c|c|c|c|c|c|c|c|c|c|c|c|c|c|c|c}
      $L$ & 0 & 1 & 2 & 3 & 4 & 5 & 6 & 7 & 8 & 9 & 10 & 11 & 12 & 13 & 14 & 
      15 & 16 & 17 & 18 & 19 & 20 & 21 & 22 & 23 & 24 & 25\\
      \hline
      \hline
      $P_L$ & $P_{\delta\delta}$ & $P_{\delta\theta}$ & $P_{\theta\theta}$ & 
      $\Pbisj{22}$ & $\Pbisj{21}$ & $\Pbisj{41}$ & $\Pbisj{40}$ & $\Pbisj{60}$ & 
      $\Ptrij{22}$ & $\Ptrij{21}$ & $\Ptrij{20}$ & $\Ptrij{42}$ & $\Ptrij{41}$ &
      $\Ptrij{40}$ & $\Ptrij{61}$ & $\Ptrij{60}$ & $\Ptrij{80}$ & 
      $P_{\delta^2\delta}$ & $P_{\delta^2\theta}$ & $P_{s^2\delta}$ & $P_{s^2\theta}$ &
      $P_{\delta^2\delta^2}$ & $P_{\delta^2s^2}$ & $P_{s^2s^2}$ & $P_{3\mathrm{nl}}$ &
      $f P_{3\mathrm{nl}}$ \\
      \hline
      $B_L$ & $b_{\delta}^2$ & $2b_\delta b_v$ & $b_v^2$ & $b_\delta^2b_v$ &
      $b_\delta b_v^2$ & $b_\delta b_v^2$ & $b_v^3$ & $b_v^3$ &
      $b_\delta^2 b_v^2$ & $b_\delta b_v^3$ & $b_v^4$ &
      $b_\delta^2 b_v^2$ & $b_\delta b_v^3$ & $b_v^4$ &
      $b_\delta b_v^3$ & $b_v^4$ & $b_v^4$ & $2b_\delta b_{\delta^2}$ &
      $2b_v b_{\delta^2}$ & $2b_\delta b_{s^2}$ & $2 b_v b_{s^2}$ &
      $b_{\delta^2}^2$ & $2 b_{\delta^2}b_{s^2}$ & $b_{s^2}^2$ &
      $2b_\delta b_{3\mathrm{nl}}$ & $2b_v b_{3\mathrm{nl}}$\\
      \hline
      $n_L$ & 0 & 2 & 4 & 2 & 2 & 4 & 4 & 6 & 2 & 2 & 2 & 4 & 4 & 4 & 
      6 & 6 & 8 & 0 & 2 & 0 & 2 & 0 & 0 & 0 & 0 & 2
    \end{tabular}
  \end{footnotesize}
  \caption{
    Power spectrum components, biases, and $\mu$ scalings.  The biased 
    redshift-space power spectrum is given by Eq.~(\ref{e:Pkmu_L}).
    \label{t:Pkmu_L}
  }
\end{table*}

Our treatment of the redshift-space power spectrum $P(k,\mu)$ includes two correction terms, $\Pbis(k,\mu)$ and $\Ptri(k,\mu)$.  A full treatment of the scale-dependent bias for these terms is beyond the scope of this paper.  Here we make the simple assumption of scale-independent density and velocity bias, with one power of $b_\delta$ for each $\delta$ index and one power of $b_v$ for each $\theta$ index.  Thus we decompose the correction terms as $\Pbis(k,\mu) = \sum_j \sum_m \mu^j b_\delta^m b_v^{3-m} \Pbisj{jm}(k)$ and $\Ptri(k,\mu) = \sum_j \sum_m \mu^j b_\delta^m b_v^{4-m} \Ptrij{jm}(k)$.  Such a decomposition allows us to write the entire redshift-space power spectrum in a massless-neutrino universe as
\begin{equation}
  P(k,\mu) 
  = 
  \Ffog(\mu\sigma_v k f)\left[
    \sum_{L=0}^{25} \mu^{n_L} B_L(\vec b) P_L(k)
    + N
    \right]
\end{equation}
with $n_L$, $B_L(\vec b)$, and $P_L(k)$ defined in Table~\ref{t:Pkmu_L}.

In the case of massive neutrinos, Ref.~\cite{Villaescusa-Navarro_2014} points out that defining the galaxy bias relative to the total matter power spectrum, rather than to the CDM+baryon power spectrum, introduces a spurious scale-dependence to the bias associated with the neutrino free-streaming scale.  Thus we define bias with respect to the CDM+baryon power spectrum:
\begin{equation}
  P(k,\mu) 
  =
  \Ffog \cdot \!
  \left[
    \sum_{L=0}^{25} \mu^{n_L} \fcb^2 \!B_L\!(\vec b) P_L
    \!+\!\! \sum_{n=0}^4 \mu^n \Peffnun
    \!\!+\!\! N \!
    \right].
  \label{e:Pkmu_L}
\end{equation}
This is the power spectrum which we compare with the data in Sec.~\ref{sec:results_and_discussion}.

\section{Time-RG with FAST-PT}
\label{sec:time-rg_with_fast-pt}

Time-Renormalization Group perturbation theory was proposed by Ref.~\cite{Pietroni_2008} and generalized to redshift space in Ref.~\cite{Upadhye_2016}.  This article was made possible by the FAST-PT techniques of Ref.~\cite{McEwen_2016}, which use Fast Fourier Transforms to compute perturbation theory integrals.  More thorough descriptions of Time-RG and FAST-PT can be found in those references.  Here we briefly describe our application of FAST-PT to the \redtime~Time-RG code of Ref.~\cite{Upadhye_2016}; note that our definitions differ slightly from that reference.

\begin{table}[tb]
  \tabcolsep=0.05cm
  \begin{footnotesize}
    \begin{tabular}{r|p{2.7in}}
      $\frac{4\pi}{k} A_{001,000}$
      &
      $
      \frac{1}{6} J_{2,2,-2}^{0001}
      +\frac{3}{4} J_{1,1,-1}^{0001}
      +\frac{1}{4} J_{0,0,0}^{0001}
      +\frac{1}{12} J_{0,2,-2}^{0001}
      +\frac{1}{6} J_{2,0,0}^{0100}
      +\frac{1}{4} J_{1,1,-1}^{0100}
      +\frac{1}{3} J_{0,0,0}^{0100}
      -\frac{1}{12} Z_{0}^{0001}
      +(Z_{-3}^{0100}
      -Z_{-1}^{0100}
      +Z_{0}^{0100}
      +\frac{1}{2} Z_{1}^{0100}
      -Z_{3}^{0001}
      +Z_{1}^{0001}
      +3 Z_{0}^{0001}
      -\frac{1}{2} Z_{-1}^{0001})/16
      $
      \\

      \hline
      $\frac{4\pi}{k} A_{001,001}$
      &
      $
      \frac{1}{6} J_{2,2,-2}^{0011}
      +\frac{1}{2} J_{1,1,-1}^{0011}
      +\frac{1}{4} J_{0,0,0}^{0011}
      +\frac{1}{12} J_{0,2,-2}^{0011}
      +\frac{1}{6} J_{2,0,0}^{0101}
      +\frac{1}{4} J_{1,1,-1}^{0101}
      +\frac{1}{4} J_{1,1,-1}^{0101}
      +\frac{1}{3} J_{0,0,0}^{0101}
      $
      \\

      \hline
      $\frac{4\pi}{k} A_{001,010}$
      &
      $
      \frac{1}{6} J_{2,2,-2}^{0101}
      +\frac{1}{2} J_{1,1,-1}^{0101}
      +\frac{1}{4} J_{0,0,0}^{0101}
      +\frac{1}{12} J_{0,2,-2}^{0101}
      +\frac{1}{6} J_{2,0,0}^{1100}
      +\frac{1}{4} J_{1,1,-1}^{1100}
      +\frac{1}{4} J_{1,1,-1}^{0011}
      +\frac{1}{3} J_{0,0,0}^{1100}
      -\frac{1}{12} Z_{0}^{0101}
      +(Z_{-3}^{1100}
      -Z_{-1}^{1100}
      +Z_{0}^{1100}
      +\frac{1}{2} Z_{1}^{1100}
      -Z_{3}^{0101}
      +Z_{1}^{0101}
      + 3 Z_{0}^{0101}
      -\frac{1}{2} Z_{-1}^{0101})/16
      $
      \\

      \hline
      $\frac{4\pi}{k} A_{001,011}$
      &
      $
      \frac{1}{6} J_{2,2,-2}^{0111}
      +\frac{3}{4} J_{1,1,-1}^{0111}
      +\frac{1}{4} J_{0,0,0}^{0111}
      +\frac{1}{12} J_{0,2,-2}^{0111}
      +\frac{1}{6} J_{2,0,0}^{1101}
      +\frac{1}{4} J_{1,1,-1}^{1101}
      +\frac{1}{3} J_{0,0,0}^{1101}
      $
      \\

      \hline
      $\frac{4\pi}{k} A_{001,100}$
      &
      $
      \frac{1}{5} J_{3,1,-1}^{0101}
      +\frac{1}{2} J_{2,0,0}^{0101}
      +\frac{1}{6} J_{2,2,-2}^{0101}
      +\frac{11}{20} J_{1,1,-1}^{0101}
      +\frac{1}{4} J_{1,1,-1}^{0101}
      +\frac{1}{4} J_{0,0,0}^{0101}
      +\frac{1}{12} J_{0,2,-2}^{0101}
      -\frac{1}{12} Z_{0}^{0011}
      +(Z_{-3}^{0101}
      -Z_{-1}^{0101}
      +Z_{0}^{0101}
      +\frac{1}{2} Z_{1}^{0101}
      -Z_{3}^{0011}
      +Z_{1}^{0011}
      + 3 Z_{0}^{0011}
      -\frac{1}{2} Z_{-1}^{0011})/16
      $
      \\

      \hline
      $\frac{4\pi}{k} A_{001,101}$
      &
      $
      \frac{1}{5} J_{3,1,-1}^{0111}
      +\frac{1}{2} J_{2,0,0}^{0111}
      +\frac{1}{6} J_{2,2,-2}^{0111}
      +\frac{11}{20} J_{1,1,-1}^{0111}
      +\frac{1}{4} J_{1,1,-1}^{1101}
      +\frac{1}{4} J_{0,0,0}^{0111}
      +\frac{1}{12} J_{0,2,-2}^{0111}
      $
      \\
    
      \hline
      $\frac{4\pi}{k} A_{001,110}$
      &
      $
      \frac{1}{5} J_{3,1,-1}^{1101}
      +\frac{1}{2} J_{2,0,0}^{1101}
      +\frac{1}{6} J_{2,2,-2}^{1101}
      +\frac{11}{20} J_{1,1,-1}^{1101}
      +\frac{1}{4} J_{1,1,-1}^{0111}
      +\frac{1}{4} J_{0,0,0}^{1101}
      +\frac{1}{12} J_{0,2,-2}^{1101}
      -\frac{1}{12} Z_{0}^{0111}
      +(Z_{-3}^{1101}
      -Z_{-1}^{1101}
      +Z_{0}^{1101}
      +\frac{1}{2} Z_{1}^{1101}
      -Z_{3}^{0111}
      +Z_{1}^{0111}
      +3 Z_{0}^{0111}
      -\frac{1}{2} Z_{-1}^{0111})/16
      $
      \\    
    
      \hline
      $\frac{4\pi}{k} A_{001,111}$
      &
      $
      \frac{1}{5} J_{3,1,-1}^{1111}
      +\frac{1}{2} J_{2,0,0}^{1111}
      +\frac{1}{6} J_{2,2,-2}^{1111}
      +\frac{11}{20} J_{1,1,-1}^{1111}
      +\frac{1}{4} J_{1,1,-1}^{1111}
      +\frac{1}{4} J_{0,0,0}^{1111}
      +\frac{1}{12} J_{0,2,-2}^{1111}
      $
      \\
    
      \hline
      $\frac{4\pi}{k} A_{111,000}$
      &
      $
      \frac{2}{5} J_{3,1,-1}^{0001}
      + J_{2,0,0}^{0001}
      +\frac{1}{3} J_{2,2,-2}^{0001}
      +\frac{11}{10} J_{1,1,-1}^{0001}
      +\frac{1}{2} J_{1,1,-1}^{0100}
      +\frac{1}{2} J_{0,0,0}^{0001}
      +\frac{1}{6} J_{0,2,-2}^{0001}
      +
      (-2 Z_{-3}^{0001}
      +2 Z_{-1}^{0001}
      -2 Z_{0}^{0001}
      -Z_{1}^{0001}
      +2 Z_{-5}^{0100}
      -4Z_{-3}^{0100}
      +Z_{-1}^{0100})/16
      $
      \\
    
      \hline
      $\frac{4\pi}{k} A_{111,001}$
      &
      $
      \frac{1}{5} J_{3,1,-1}^{0011}
      +\frac{1}{2} J_{2,0,0}^{0011}
      +\frac{1}{6} J_{2,2,-2}^{0011}
      +\frac{11}{20} J_{1,1,-1}^{0011}
      +\frac{1}{4} J_{1,1,-1}^{1100}
      +\frac{1}{4} J_{0,0,0}^{0011}
      +\frac{1}{12} J_{0,2,-2}^{0011}
      +\frac{1}{5} J_{3,1,-1}^{0101}
      +\frac{1}{2} J_{2,0,0}^{0101}
      +\frac{1}{6} J_{2,2,-2}^{0101}
      +\frac{11}{20} J_{1,1,-1}^{0101}
      +\frac{1}{4} J_{1,1,-1}^{0101}
      +\frac{1}{4} J_{0,0,0}^{0101}
      +\frac{1}{12} J_{0,2,-2}^{0101}
      +
      (-Z_{-3}^{0101}
      +Z_{-1}^{0101}
      -Z_{0}^{0101}
      -\frac{1}{2} Z_{1}^{0101}
      +Z_{-5}^{1100}
      - 2 Z_{-3}^{1100}
      +\frac{1}{2} Z_{-1}^{1100})/16
      $
      \\
    
      \hline
      $\frac{4\pi}{k} A_{111,011}$
      &
      $
      \frac{2}{5} J_{3,1,-1}^{0111}
      + J_{2,0,0}^{0111}
      +\frac{1}{3} J_{2,2,-2}^{0111}
      +\frac{11}{10} J_{1,1,-1}^{0111}
      +\frac{1}{2} J_{1,1,-1}^{1101}
      +\frac{1}{2} J_{0,0,0}^{0111}
      +\frac{1}{6} J_{0,2,-2}^{0111}
      $
      \\
    
      \hline
      $\frac{4\pi}{k} A_{111,100}$
      &
      $
      \frac{8}{35} J_{4,0,0}^{0101}
      +\frac{4}{5} J_{3,1,-1}^{0101}
      +\frac{19}{21} J_{2,0,0}^{0101}
      +\frac{1}{3} J_{2,2,-2}^{0101}
      +\frac{6}{5} J_{1,1,-1}^{0101}
      +\frac{11}{30} J_{0,0,0}^{0101}
      +\frac{1}{6} J_{0,2,-2}^{0101}
      +
      (-2 Z_{-3}^{0011}
      +2 Z_{-1}^{0011}
      -2 Z_{0}^{0011}
      -Z_{1}^{0011}
      +2 Z_{-5}^{0101}
      -4 Z_{-3}^{0101}
      +Z_{-1}^{0101})/16
      $
      \\
    
      \hline
      $\frac{4\pi}{k} A_{111,101}$
      &
      $
      \frac{8}{35} J_{4,0,0}^{0111}
      +\frac{2}{5} J_{3,1,-1}^{0111}
      +\frac{2}{5} J_{3,1,-1}^{1101}
      +\frac{19}{21} J_{2,0,0}^{0111}
      +\frac{1}{6} J_{2,2,-2}^{0111}/6
      +\frac{1}{6} J_{2,2,-2}^{1101}/6
      +\frac{3}{5} J_{1,1,-1}^{0111}
      +\frac{3}{5} J_{1,1,-1}^{1101}
      +\frac{11}{30} J_{0,0,0}^{0111}
      +\frac{1}{12} J_{0,2,-2}^{0111}
      +\frac{1}{12} J_{0,2,-2}^{1101}
      +
      (-Z_{-3}^{0111}
      +Z_{-1}^{0111}
      -Z_{0}^{0111}
      -\frac{1}{2} Z_{1}^{0111}
      +Z_{-5}^{1101}
      -2 Z_{-3}^{1101}
      +\frac{1}{2} Z_{-1}^{1101})/16
      $
      \\
    
      \hline
      $\frac{4\pi}{k} A_{111,111}$
      &
      $
      \frac{8}{35} J_{4,0,0}^{1111}
      +\frac{4}{5} J_{3,1,-1}^{1111}
      +\frac{19}{21} J_{2,0,0}^{1111}
      +\frac{1}{3} J_{2,2,-2}^{1111}
      +\frac{6}{6} J_{1,1,-1}^{1111}
      +\frac{11}{30} J_{0,0,0}^{1111}
      +\frac{1}{6} J_{0,2,-2}^{1111}
      $
      \\
    \end{tabular}
  \end{footnotesize}
  \caption{
    FAST-PT expansions of $A_{acd,bef}(k)$.
    All other non-zero $A_{acd,bef}$ are related to the above by the
    identity $A_{adc,bfe}(k) = A_{acd,bef}(k)$.
    \label{t:redTime_fastpt_Aacdbef}
  }
\end{table}

\begin{table}[tb]
  \tabcolsep=0.05cm
  \begin{footnotesize}
    \begin{tabular}{r|p{2.9in}}
      $2\pi k R^{(1)}_{abc}$
      &
      $
      \dk_{a0}[
        \frac{2}{5} J_{3,1,-1}^{0bc1}
        -\frac{7}{5} J_{1,1,-1}^{0bc1}
        - J_{1,1,-1}^{1c0b}
        - 2 J_{0,0,0}^{0bc1}
        +\frac{2}{5} J_{3,1,-1}^{0cb1}
        +\frac{2}{3} J_{2,0,0}^{1b0c}
        -\frac{2}{3} J_{2,2,-2}^{0cb1}
        -\frac{12}{5} J_{1,1,-1}^{0cb1}
        -\frac{5}{3} J_{0,0,0}^{1b0c}
        -\frac{1}{3} J_{0,2,-2}^{0cb1}
        ]
      + \dk_{b0}[
        -\frac{13}{12} Z_{0}^{0ca1}
        +\frac{5}{16} Z_{-1}^{0ca1}
        -\frac{7}{16} Z_{1}^{0ca1}
        -\frac{1}{8}Z_{-3}^{0ca1}
        +\frac{3}{8}Z_{3}^{0ca1}
        -\frac{3}{8}Z_{0}^{1c0a}
        +\frac{7}{16} Z_{-1}^{1c0a}
        -\frac{3}{16} Z_{1}^{1c0a}
        -\frac{5}{8}Z_{-3}^{1c0a}
        +\frac{1}{8}Z_{-5}^{1c0a}
        ]
      + \dk_{c0}[
        \frac{1}{8}Z_{-5}^{1b0a}
        -\frac{3}{8}Z_{-3}^{1b0a}
        +\frac{3}{16} Z_{-1}^{1b0a}
        -\frac{1}{16} Z_{1}^{1b0a}
        -\frac{1}{8}Z_{0}^{1b0a}
        -\frac{1}{8}Z_{-3}^{0ba1}
        +\frac{3}{16} Z_{-1}^{0ba1}
        -\frac{3}{16} Z_{1}^{0ba1}
        +\frac{1}{8}Z_{3}^{0ba1}
        ]
      + \dk_{a1}[
        \frac{16}{35} J_{4,0,0}^{b1c1}
        -\frac{2}{5} J_{3,1,-1}^{c1b1}
        +\frac{2}{5} J_{3,1,-1}^{b1c1}
        -\frac{46}{21} J_{2,0,0}^{b1c1}
        -\frac{2}{3} J_{2,2,-2}^{b1c1}
        -\frac{13}{5} J_{1,1,-1}^{c1b1}
        -\frac{7}{5} J_{1,1,-1}^{b1c1}
        -\frac{19}{15} J_{0,0,0}^{b1c1}
        -\frac{1}{3} J_{0,2,-2}^{c1b1}
        ]
      + \dk_{b1}[
        -\frac{1}{3} Z_{0}^{c1a1}
        ]
      + \dk_{c1}[
        \frac{1}{3} Z_{0}^{b1a1}
        ]
      $
      \\
    
      \hline
      $2\pi k R^{(2)}_{abc}$
      &
      $
      \dk_{a0}[
        \frac{3}{5}J_{3,1,-1}^{0bc1}
        +J_{2,0,0}^{0bc1}
        -\frac{3}{5}J_{1,1,-1}^{0bc1}
        -J_{0,0,0}^{0bc1}
        +\frac{3}{5}J_{3,1,-1}^{0cb1}
        +J_{2,0,0}^{1b0c}
        -\frac{3}{5}J_{1,1,-1}^{0cb1}
        -J_{0,0,0}^{1b0c}
        ]
      + \dk_{b0}[
        -\frac{1}{2} Z_{0}^{0ca1}
        +\frac{9}{32} Z_{-1}^{0ca1}
        -\frac{9}{32} Z_{1}^{0ca1}
        -\frac{3}{16} Z_{-3}^{0ca1}
        +\frac{3}{16} Z_{3}^{0ca1}
        -\frac{3}{16} Z_{0}^{1c0a}
        -\frac{3}{32} Z_{1}^{1c0a}
        +\frac{9}{32} Z_{-1}^{1c0a}
        -\frac{9}{16} Z_{-3}^{1c0a}
        +\frac{3}{16} Z_{-5}^{1c0a}
        ]
      + \dk_{c0}[
        \frac{3}{16} Z_{-5}^{1b0a}
        -\frac{9}{16} Z_{-3}^{1b0a}
        +\frac{9}{32} Z_{-1}^{1b0a}
        -\frac{3}{32} Z_{1}^{1b0a}
        -\frac{3}{16} Z_{0}^{1b0a}
        +\frac{3}{16} Z_{3}^{0ba1}
        -\frac{3}{16} Z_{-3}^{0ba1}
        -\frac{9}{32} Z_{1}^{0ba1}
        +\frac{9}{32} Z_{-1}^{0ba1}
        -\frac{1}{2} Z_{0}^{0ba1}
        ]
      + \dk_{a1}[
        \frac{24.}{35} J_{4,0,0}^{b1c1}
        - J_{3,1,-1}^{c1b1}
        +\frac{11}{5}J_{3,1,-1}^{b1c1}
        -\frac{2}{7} J_{2,0,0}^{b1c1}
        -\frac{3}{5}J_{1,1,-1}^{b1c1}
        -\frac{3}{5}J_{1,1,-1}^{c1b1}
        -\frac{2}{5}J_{0,0,0}^{b1c1}
        ]
      $
      \\
    
      \hline
      $2\pi k R^{(3)}_{abc}$
      &
      $
      \dk_{a0}[
        (
        \frac{4}{7} J_{4,0,2}^{1c0b}
        -\frac{40}{21} J_{2,0,2}^{1c0b}
        +\frac{4}{3} J_{0,0,2}^{1c0b}
        -\frac{4}{7} J_{4,0,2}^{1b0c}
        +\frac{40}{21} J_{2,0,2}^{1b0c}
        -\frac{4}{3} J_{0,0,2}^{1b0c}
        )/k^2
        -J_{3,1,-1}^{0bc1}
        +J_{1,1,-1}^{0bc1}
        -\frac{5}{3} J_{2,0,0}^{1b0c}
        +\frac{5}{3} J_{0,0,0}^{1b0c}
        ]
      + \dk_{b0}[
        \frac{35}{32} Z_{0}^{0ca1}
        +\frac{5}{32} Z_{5}^{0ca1}
        -\frac{5}{8} Z_{3}^{0ca1}
        +\frac{5}{32} Z_{-3}^{0ca1}
        -\frac{5}{16} Z_{-1}^{0ca1}
        +\frac{15}{32} Z_{1}^{0ca1}
        +\frac{55}{96} Z_{0}^{1c0a}
        -\frac{5}{32} Z_{-5}^{1c0a}
        +\frac{5}{8} Z_{-3}^{1c0a}
        -\frac{5}{32} Z_{3}^{1c0a}
        -\frac{15}{32} Z_{-1}^{1c0a}
        +\frac{5}{16} Z_{1}^{1c0a}
        ]
      + \dk_{c0}[
        -\frac{5}{32}Z_{-5}^{1b0a}
        +\frac{5}{16}Z_{-3}^{1b0a}
        -\frac{25}{96} Z_{0}^{1b0a}
        -\frac{5}{32}Z_{1}^{1b0a}
        +\frac{5}{32}Z_{3}^{1b0a}
        -\frac{5}{32}Z_{5}^{0ba1}
        +\frac{5}{16}Z_{3}^{0ba1}
        -\frac{25}{96} Z_{0}^{0ba1}
        -\frac{5}{32}Z_{-1}^{0ba1}
        +\frac{5}{32}Z_{-3}^{0ba1}
        ]
      + \dk_{a1}[
        -\frac{4}{7} J_{4,0,0}^{b1c1}
        -J_{3,1,-1}^{b1c1}
        +\frac{5}{21} J_{2,0,0}^{b1c1}
        +J_{1,1,-1}^{b1c1}
        +\frac{1}{3} J_{0,0,0}^{b1c1}
        ]
      + \dk_{b1}[
        \frac{1}{3} Z_{0}^{c1a1}
        ]
      + \dk_{c1}[
        -\frac{1}{3} Z_{0}^{b1a1}
        ]
      $
      \\
    \end{tabular}
  \end{footnotesize}
  \caption{
    FAST-PT expansions of $\Rlabc(k)$. $\dk_{ab}$ is the Kronecker delta 
    function.
    \label{t:redTime_fastpt_Rlabc}
  }
\end{table}

\begin{table}[tb]
  \tabcolsep=0.1cm
  \begin{footnotesize}
    \begin{tabular}{r|p{2.8in}}
      $\Ptrij{22}$
      &
      $
      \frac{1}{3} J_{2,0,0}^{0101}
      -\frac{1}{3} J_{0,0,0}^{0101}
      $
      \\

      \hline
      $\Ptrij{21}$
      &
      $
      (
      -\frac{6}{35} J_{4,0,2}^{1101}
      +\frac{4}{7} J_{2,0,2}^{1101}
      -\frac{2}{5} J_{0,0,2}^{1101}
      )/k^2
      $
      \\
      
      \hline
      $\Ptrij{20}$
      &
      $
      (\frac{5}{231} J_{6,2,2}^{1111}
      -\frac{9}{77} J_{4,2,2}^{1111}
      +\frac{5}{21} J_{2,2,2}^{1111}
      -\frac{1}{7} J_{0,2,2}^{1111}
      )/k^4
      $
      \\
      
      \hline
      $\Ptrij{42}$
      &
      $
      \frac{1}{3} J_{2,0,0}^{0101}
      +2 J_{1,1,-1}^{0101}
      +\frac{5}{3} J_{0,0,0}^{0101}
      $
      \\
      
      \hline
      $\Ptrij{41}$
      &
      $
      -\frac{6}{5} J_{3,1,-1}^{0111}
      +2 J_{2,0,0}^{1101}
      +\frac{6}{5} J_{1,1,-1}^{0111}
      -2 J_{0,0,0}^{1101}
      +(
      \frac{12}{7} J_{4,0,2}^{1101}
      -\frac{40}{7} J_{2,0,2}^{1101}
      +4 J_{0,0,2}^{1101}
      )/k^2
      $
      \\
      
      \hline
      $\Ptrij{40}$
      &
      $
      (-\frac{5}{11} J_{6,2,2}^{1111}
      +\frac{27}{11} J_{4,2,2}^{1111}
      -5 J_{2,2,2}^{1111}
      +3 J_{0,2,2}^{1111}
      )/k^4
      +(
      -\frac{9}{7} J_{4,0,2}^{1111}
      +\frac{30}{7} J_{2,0,2}^{1111}
      -3 J_{0,0,2}^{1111}
      )/k^2
      +\frac{27}{70} J_{4,0,0}^{1111}
      -\frac{9}{7} J_{2,0,0}^{1111}
      +\frac{9}{10} J_{0,0,0}^{1111}
      $
      \\
      
      \hline
      $\Ptrij{61}$
      &
      $
      (-2 J_{4,0,2}^{1101}
      +\frac{20}{3} J_{2,0,2}^{1101}
      -\frac{14}{3} J_{0,0,2}^{1101}
      )/k^2
      +2 J_{3,1,-1}^{0111}
      -\frac{2}{3} J_{2,0,0}^{1101}
      +2 J_{1,1,-1}^{1101}
      +\frac{14}{3} J_{0,0,0}^{1101}
      $
      \\
      
      \hline
      $\Ptrij{60}$
      &
      $
      (\frac{15}{11} J_{6,2,2}^{1111}
      -\frac{81}{11} J_{4,2,2}^{1111}
      +15 J_{2,2,2}^{1111}
      -9 J_{0,2,2}^{1111}
      )/k^4
      +(6 J_{4,0,2}^{1111}
      -20 J_{2,0,2}^{1111}
      +14 J_{0,0,2}^{1111}
      )/k^2
      -\frac{39}{35} J_{4,0,0}^{1111}
      -\frac{6}{5} J_{3,1,-1}^{1111}
      +\frac{47}{7} J_{2,0,0}^{1111}
      +\frac{6}{5} J_{1,1,-1}^{1111}
      -\frac{28}{5} J_{0,0,0}^{1111}
      $
      \\
      
      \hline
      $\Ptrij{80}$
      &
      $
      (- J_{6,2,2}^{1111}
      +\frac{27}{5} J_{4,2,2}^{1111}
      -11 J_{2,2,2}^{1111}
      +\frac{33}{5} J_{0,2,2}^{1111}
      )/k^4
      +(
      -\frac{27}{5} J_{4,0,2}^{1111}
      +18 J_{2,0,2}^{1111}
      -\frac{63}{5} J_{0,0,2}^{1111}
      )/k^2
      +\frac{59}{70} J_{4,0,0}^{1111}
      +2 J_{3,1,-1}^{1111}
      -\frac{36}{7} J_{2,0,0}^{1111}
      +\frac{63}{10} J_{0,0,0}^{1111}
      $
      \\
      
    \end{tabular}
  \end{footnotesize}
  \caption{
    FAST-PT expansions of $P^{\mathrm T}_{jm}(k)$.
    \label{t:redTime_fastpt_PTjm}
  }
\end{table}

\begin{table}[tb]
  \tabcolsep=0.1cm
  \begin{footnotesize}
    \begin{tabular}{r|p{2.8in}}
      $P_{\delta^2,\delta}$
      &
      $
      \frac{4}{21} J_{2,0,0}^{0000}
      +J_{1,1,-1}^{0000}
      +\frac{17}{21} J_{0,0,0}^{0000}
      $
      \\

      \hline
      $P_{\delta^2,\theta}$
      &
      $
      \frac{8}{21} J_{2,0,0}^{0000}
      +J_{1,1,-1}^{0000}
      +\frac{13}{21} J_{0,0,0}^{0000}
      $
      \\

      \hline
      $P_{s^2,\delta}$
      &
      $
      \frac{16}{245} J_{4,0,0}^{0000}
      +\frac{2}{5} J_{3,1,-1}^{0000}
      +\frac{254}{441} J_{2,0,0}^{0000}
      +\frac{4}{15} J_{1,1,-1}^{0000}
      +\frac{8}{315} J_{0,0,0}^{0000}
      $
      \\

      \hline
      $P_{s^2,\theta}$
      &
      $
      \frac{32}{245} J_{4,0,0}^{0000}
      +\frac{2}{5} J_{3,1,-1}^{0000}
      +\frac{214}{441} J_{2,0,0}^{0000}
      +\frac{4}{15} J_{1,1,-1}^{0000}
      +\frac{16}{315} J_{0,0,0}^{0000}
      $
      \\

      \hline
      $P_{\delta^2,\delta^2}$
      &
      $
      \frac{1}{2}J_{0,0,0}^{0000}-\frac{1}{2}J_{0,0,0}^{0000}(0)
      $
      \\

      \hline
      $P_{\delta^2,s^2}$
      &
      $
      \frac{1}{3} J_{2,0,0}^{0000} - \frac{1}{3}J_{0,0,0}^{0000}(0)
      $
      \\

      \hline
      $P_{s^2,s^2}$
      &
      $
      \frac{4}{35} J_{4,0,0}^{0000}
      +\frac{4}{63} J_{2,0,0}^{0000}
      +\frac{2}{45} J_{0,0,0}^{0000}
      -\frac{2}{9} J_{0,0,0}^{0000}(0)
      $
      \\

      \hline
      $P_{3\mathrm{nl}}$
      &
      $
      -\frac{15}{256} Z_{-5}^{0000}
      +\frac{15}{64} Z_{-3}^{0000}
      -\frac{15}{256} Z_{3}^{0000}
      -\frac{45}{256} Z_{-1}^{0000}
      +\frac{15}{128} Z_{1}^{0000}
      +\frac{55}{256} Z_{0}^{0000}
      $
      \\
    \end{tabular}
  \end{footnotesize}
  \caption{
    FAST-PT expansions of the scale-dependent bias terms of
    McDonald and Roy, Ref.~\cite{McDonald_Roy_2009}. The $k$-dependence of
    the bias terms, the $J_{\ell\alpha\beta}^{abcd}(k)$, and the
    $Z_{\mathcal N}^{abcd}(k)$ has been suppressed, except for $J_{0,0,0}^{0000}(0)$,
    the low-$k$ limit of $J_{0,0,0}^{0000}(k)$.
    \label{t:redTime_fastpt_PMR}
  }
\end{table}


Define $\eta = \log(\frac{1+z_\mathrm{in}}{1+z})$ for initial redshift $z_\mathrm{in}$, as well as $\varphi_0 = e^{-\eta}\delta$ and $\varphi_1 = e^{-\eta}\theta$.  Then the continuity and Euler equations in Fourier space are
\begin{eqnarray}
  \varphi_a' + \Xi_{ab} \varphi_b
  &=&
  e^\eta \int \frac{d^3q d^3p}{(2\pi)^3} \Dirac(\vec k - \vec q - \vec p)
  \gamma_{abc}^{\vec k,\vec q, \vec p} \varphi_b^{\vec q} \varphi_c^{\vec p} \qquad
  \\
  \Xi_{00}
  &=&
  -\Xi_{01} = 1
  \\
  \Xi_{10}
  &=&
  -\frac{3}{2} \Omegam(\eta) \left(\fcb + \fnu\frac{\deltanu}{\deltacb}\right)
  \\
  \Xi_{11}
  &=&
  2 + \Hc'/\Hc
  \\
  \gamma_{001}^{\vec k, \vec q, \vec p}
  &=&
  \gamma_{010}^{\vec k, \vec p, \vec q}
  =
  (\vec q + \vec p)\cdot \vec p / (2p^2)
  \\
  \gamma_{111}^{\vec k,\vec q,\vec p}
  &=&
  (\vec q + \vec p)^2 \vec q \cdot \vec p / (2 q^2 p^2)
\end{eqnarray}
where primes denote $\partial / \partial\eta$, summation over repeated indices is implicit, and vectors in superscripts are shorthand for arguments; for example, $\varphi_i^{\vec q} = \varphi_i(\vec q)$. The matter fraction is $\Omegam(\eta) = \Omegamo H_0^2 (1+z)^3 / H^2$.

Evolution equations  of coordinate-space Time-RG are
\begin{eqnarray}
  P_{ab}'
  &=&
  \!-\!\Xi_{ac}P_{bc} \!-\! \Xi_{bc}P_{ac}
  \!\!+\!\! \frac{4\pi e^\eta}{k} \!(\!I_{acd,bcd} \!+\! I_{bcd,acd}\!)\qquad
  \\
  \frac{4\pi}{k} I_{acd,bef} \!\!
  &=&
  \!\!\int\!\! \frac{d^3q}{(2\pi)^3} \gamma_{acd}^{\vec k,\vec q, \vec p}
  B_{bef}^{\vec k,\vec q,\vec p}
  \\
  I_{acd,bef}'\!\!
  &=&
  -\Xi_{bg}I_{acd,gef} - \Xi_{eq}I_{acd,bgf} - \Xi_{fg}I_{acd,beg}
  \nonumber\\
  &~& + 2e^\eta A_{acd,bef}
  \\
  \frac{4\pi}{k} A_{acd,bef} \!\!
  &=&
  \!\!\int\!\! \frac{d^3q}{(2\pi)^3} \gamma_{acd}^{\vec k,\vec q,\vec p}
  (\gamma_{bgh}^{\vec k,\vec q,\vec p} P_{ge}^{\vec q} P_{hf}^{\vec p}
  \! +\! \gamma_{egh}^{\vec q,\vec p,\vec k} P_{gf}^{\vec p} P_{hb}^{\vec k}
  \nonumber\\
  &~&\qquad\qquad\qquad
  + \gamma_{fgh}^{\vec p,\vec k,\vec q} P_{gb}^{\vec k} P_{he}^{\vec q})
\end{eqnarray}
where $\vec p = \vec k - \vec q$, and the $k$- and $\eta$-dependence of $P_{ab}$, $I_{acd,bef}$, and $A_{acd,bef}$ have been suppressed.  These are initialized at $z_\mathrm{in}$ sufficiently large that $P_{ab}$ is linear and the bispectrum is negligible; we choose $z_\mathrm{in}=200$.  Redshift-space Time-RG similarly decomposes the bispectrum-dependence of $\Pbisj{jm}$ as
\begin{eqnarray}
  \Pbisj{22}
  &=&
  -2\pi k \QQ{1}{010} + \frac{4\pi k}{3} \QQ{2}{010}
  \\
  \Pbisj{21}
  &=&
  \frac{4\pi k}{3} \QQ{2}{011} + \frac{6\pi k}{5} \QQ{3}{011}
  \\
  \Pbisj{41}
  &=&
  -2\pi k \QQ{1}{110} \!+\! \frac{4\pi k}{3}\QQ{2}{110}
  \!-\! 2\pi k \QQ{1}{011} \!-\! 2\pi k \QQ{3}{011} \qquad
  \\
  \Pbisj{40}
  &=&
  \frac{4\pi k}{3} \QQ{2}{111} + \frac{6\pi k}{5} \QQ{3}{111}
  \\
  \Pbisj{60}
  &=&
  -2\pi k \QQ{1}{111} - 2\pi k \QQ{3}{111}
\end{eqnarray}
where the functions $\Qlabc(k)$ are defined as
\begin{eqnarray}
  \pi \QQ{1}{abc}
  &=&
  \!\!\int\!\! \frac{d^3q}{(2\pi)^3} \frac{k}{p^2} B_{abc}^{\vec k,\vec q,\vec p}
  \times
  \nonumber\\
  &~&
  \left[ 2 \Pleg_2(\hat q \!\cdot\! \hat k)
    \!+\! (\tfrac{q}{k} \!+\! \tfrac{k}{q} )
    \Pleg_1(\hat q \!\cdot\! \hat k) \right]\qquad
  \\
  \pi \QQ{2}{abc}
  &=&
  \!\!\int\!\! \frac{d^3q}{(2\pi)^3} \frac{k}{p^2} B_{abc}^{\vec k,\vec q,\vec p}
  \!\left[ \Pleg_2(\hat q \!\cdot\! \hat k)
    - \Pleg_0(\hat q \!\cdot\! \hat k) \right]\qquad
  \\
  \pi \QQ{3}{abc}
  &=&
  \!\!\int\!\! \frac{d^3q}{(2\pi)^3} \frac{q}{p^2} B_{abc}^{\vec k,\vec q,\vec p}
  \!\left[ \Pleg_3(\hat q \!\cdot\! \hat k)
    - \Pleg_1(\hat q \!\cdot\! \hat k) \right]
\end{eqnarray}
and the $\Pleg_\ell$ are Legendre polynomials.  $\Qlabc$ evolve as
\begin{eqnarray}
  {\Qlabc}'
  &=&
  \!- \Xi_{ad}\QQ{\ell}{dbc} \!-\! \Xi_{bd}\QQ{\ell}{adc}
  \!-\!\Xi_{cd}\QQ{\ell}{abd}
  \!+\! 2 e^\eta \Rlabc\qquad
  \\
  \pi R^{(1)}_{abc}
  &=&
    \!\!\int\!\! \frac{d^3q}{(2\pi)^3} \frac{k}{p^2} 
  \left[ 2 \Pleg_2(\hat q \!\cdot\! \hat k)
    \!+\! (\tfrac{q}{k} \!+\! \tfrac{k}{q} )
    \Pleg_1(\hat q \!\cdot\! \hat k) \right] \times
  \nonumber\\
  &~&
  (\gamma_{ade}^{\vec k,\vec q,\vec p}P_{db}^{\vec q} P_{ec}^{\vec p}
  \!+\! \gamma_{bde}^{\vec q,\vec p,\vec k}P_{dc}^{\vec p} P_{ea}^{\vec k}
  \!+\! \gamma_{cde}^{\vec p,\vec k,\vec q}P_{da}^{\vec k} P_{eb}^{\vec q})\qquad\quad
  \\
  \pi R^{(2)}_{abc}
  &=&
  \!\!\int\!\! \frac{d^3q}{(2\pi)^3} \frac{k}{p^2} 
  \left[
    \Pleg_2(\hat q \!\cdot\! \hat k)
    - \Pleg_0(\hat q \!\cdot\! \hat k)
    \right] \times
  \nonumber\\
  &~&
  (\gamma_{ade}^{\vec k,\vec q,\vec p}P_{db}^{\vec q} P_{ec}^{\vec p}
  \!+\! \gamma_{bde}^{\vec q,\vec p,\vec k}P_{dc}^{\vec p} P_{ea}^{\vec k}
  \!+\! \gamma_{cde}^{\vec p,\vec k,\vec q}P_{da}^{\vec k} P_{eb}^{\vec q})\qquad\quad
  \\
  \pi R^{(3)}_{abc}
  &=&
  \!\!\int\!\! \frac{d^3q}{(2\pi)^3} \frac{q}{p^2} 
  \left[
    \Pleg_3(\hat q \!\cdot\! \hat k)
    - \Pleg_1(\hat q \!\cdot\! \hat k)
    \right] \times
  \nonumber\\
  &~&
  (\gamma_{ade}^{\vec k,\vec q,\vec p}P_{db}^{\vec q} P_{ec}^{\vec p}
  \!+\! \gamma_{bde}^{\vec q,\vec p,\vec k}P_{dc}^{\vec p} P_{ea}^{\vec k}
  \!+\! \gamma_{cde}^{\vec p,\vec k,\vec q}P_{da}^{\vec k} P_{eb}^{\vec q}).
\end{eqnarray}

The FAST-PT method of Ref.~\cite{McEwen_2016} decomposes the mode-coupling and convolution integrals of perturbation theory into terms of the form
\begin{eqnarray}
  J_{\ell\alpha\beta}^{abcd}(k)
  &=&
  \!\int\! \frac{d^3q}{(2\pi)^3} q^\alpha p^\beta
  \Pleg_\ell(\hat q \!\cdot\! \hat p)
  P_{ab}(q) P_{cd}(p) \quad
  \\
  Z_{\mathcal N}^{abcd}(k)
  &=&
  \!\int\! \frac{d^3q}{(2\pi)^3} \lambda_{\mathcal N}(q/k)
  P_{ab}(q) P_{cd}(k)
\end{eqnarray}
where
\begin{eqnarray}
  \lambda_0(r)
  &=&
  1
  \\
  \lambda_1(r)
  &=&
  (1-r) \log\left|\frac{1+r}{1-r}\right|
  \\
  \lambda_2(r)
  &=&
  r + \frac{1}{2}(1-r^2) \log\left|\frac{1+r}{1-r}\right|
  \\
  \lambda_3(r)
  &=&
  r^2 + \frac{1}{2}(1-r^3) \log\left|\frac{1+r}{1-r}\right|
  \end{eqnarray}\begin{eqnarray}
  \lambda_4(r)
  &=&
  r^3 + \frac{1}{3}r + \frac{1}{2}(1-r^4) \log\left|\frac{1+r}{1-r}\right|
  \\
  \lambda_5(r)
  &=&
  r^4 + \frac{1}{3}r^2 + \frac{1}{2}(1-r^5) \log\left|\frac{1+r}{1-r}\right|
\end{eqnarray}
and $\lambda_{-\mathcal N}(r) = \lambda_{\mathcal N}(1/r)$.  That reference computes these using FFTs and shows how to regularize the divergent terms.  All that remains is to expand quantities of interest in the $ J_{\ell\alpha\beta}^{abcd}(k)$s and $Z_{\mathcal N}^{abcd}(k)$s.  This is done for $A_{acd,bef}(k)$, $\Rlabc(k)$, $\Ptrij{jm}(k)$, and the McDonald-Roy bias terms in Tables~\ref{t:redTime_fastpt_Aacdbef}, \ref{t:redTime_fastpt_Rlabc}, \ref{t:redTime_fastpt_PTjm}, and \ref{t:redTime_fastpt_PMR}, respectively.  The $1$-loop version of Time-RG used here computes $ J_{\ell\alpha\beta}^{abcd}(k)$ and $Z_{\mathcal N}^{abcd}(k)$ using the linear power spectra.

\section{BOSS likelihood}
\label{sec:boss_likelihood}

The BOSS DR11 analysis of Ref.~\cite{Beutler_2014} measures the monopole and quadrupole power spectra binned by wave numbers $k_{\tilde i}$, with $0 \leq {\tilde i} < 38$.  In order to compare a cosmological model with these data, we construct windowed multipole power spectra from Eq.~(\ref{e:Pkmu_L}). Following Ref.~\cite{Upadhye_2016} we express the multipoles of $P(k,\mu)$ as $P^{(\ell)}(k) = \sum_n {\mathcal M}_{\ell n}(f k \sigma_v)  P_{n}(k)$, where $P(k,\mu) = \Ffog(f k \sigma_v \mu) \sum_n \mu^n P_n(k)$.  The coefficients ${\mathcal M}_{\ell n}(f k \sigma_v)$ depend on $\Ffog$.  For the Lorentzian streaming functions used here, $ {\mathcal M}_{\ell n}(\alpha) = \frac{2\ell+1}{2} \sum_{n'} p_{\ell,n'} m_{n+n'}(\alpha)$, where $p_{\ell,n'}$ are the coefficients of the Legendre polynomials ${\mathscr P}_\ell(x) = \sum_{n} p_{\ell,n} x^n$, and the $m_n$ are given by the recursion relation $\alpha^2 m_n = 2/(2n-1) - m_{n-1}$, $m_0 = 2 \, \mathrm{arctan}(\alpha)/\alpha$.  Using the window functions $w_{\ell \ell'}(k_{\tilde i},q)$ of Ref.~\cite{Beutler_2014}, we write the binned, windowed model power spectra as
\begin{equation}
  P^{(\ell,{\tilde i})}
  =
  2\pi \!\int\!\! q^2 dq
  \!\!\sum_{\ell'=0,2}\!\! w_{\ell \ell'}(k_{\tilde i},q)
  \!\sum_n {\mathcal M}_{\ell' n} P_n(q).
  \label{e:binned_model_power1}
\end{equation}
Reference~\cite{Beutler_2014} measures the monopole ($\ell=0$) and quadrupole ($\ell=2$) power spectra.  Henceforth we use a shorthand notation combining $\ell$ and $\tilde i$ into a single integer $i$ ranging from $0$ to $75$, with $0 \leq i \leq 37$ corresponding to ($\ell=0, \tilde i = i$), and $38 \leq i \leq 75$ corresponding to ($\ell = 2, \tilde i = i - 38$).  Further simplification is possible by pulling bias-dependent terms out of the integral,
\begin{eqnarray}
  P_i
  &=&
  \fcb^2 \sum_{L=0}^{25} B_L\!(\vec b) I_{Li}^{(c)}
  + I_i^{(\nu)} + N I_i^{(N)}
  \label{e:binned_model_power2}
  \\
  I_{Li}^{(c)}
  &=&
  2\pi \!\!\int\!\! q^2 dq \sum_{\ell'}
  w_{\ell\ell'}(k_{\tilde i},q) {\mathcal M}_{\ell' n_L}\!(f \sigma_v q) P_\ell(q)
  \label{e:binned_Ic}
  \\
  I_i^{(\nu)}
  &=&
  2\pi \!\!\int\!\! q^2 dq \sum_{\ell', n}
  w_{\ell\ell'}(k_{\tilde i},q) {\mathcal M}_{\ell' n}\!(f \sigma_v q)
  \Peffnun\!(q) \qquad
  \label{e:binned_Inu}
  \\
  I_i^{(N)}
  &=&
  2\pi \!\!\int\!\! q^2 dq  \sum_{\ell'} {\mathcal M}_{\ell' 0}(f \sigma_v q).
  \label{e:binned_IN}
\end{eqnarray}

Our likelihood calculation for BOSS DR11 data follows the treatment of Ref.~\cite{Beutler_2014}.  Here we detail our computation, designed to facilitate marginalization over the bias parameters.  Up to a normalization constant, the likelihood of a model with cosmological parameters $\vec c$ and bias parameters $\vec b$ is ${\mathcal L} \propto \exp[-\chi(\vec c, \vec b)^2/2]$, with
\begin{equation}
  \chi^2
  =
  \sum_{i,j}
  ({\mathbf C}^{-1})_{ij}
  \left[ P^\mathrm{d}_i - P^\mathrm{t}_i(\vec c, \vec b) \right]
  \left[ P^\mathrm{d}_j - P^\mathrm{t}_j(\vec c, \vec b) \right].
  \label{e:chi2}
\end{equation}
Here, $\mathbf C$ is the covariance matrix of the BOSS DR11 data, from Ref.~\cite{Beutler_2014}; $P^\mathrm{d}_i$ is the binned BOSS power spectrum; and $P^\mathrm{t}_j(\vec c, \vec b)$ is the binned, windowed model power spectrum of Eq.~(\ref{e:binned_model_power2}).  In practice, Ref.~\cite{Beutler_2014} provides separate data sets for the northern and southern sky patches.  We compute $\chi^2$ as in Eq.~(\ref{e:chi2}) for each patch and then sum them to find the total $\chi^2$.

Once again, we  pull bias-dependent factors outside the summations:
\begin{eqnarray}
  \chi^2
  &=&
  \fcb^4 B_L B_M \xcc_{LM} + 2 \fcb^2 B_L N \xcN_L 
  \nonumber\\
  &&
  + 2\fcb^2 B_L \xcnu_L + N^2 \xNN + 2 N \xnuN + \xnunu
  \nonumber\\
  &&
  + \xdd \!-2\fcb^2 B_L \xdc_L \!- 2 N \xdN - 2\xdnu
  \label{e:chiSq}
\end{eqnarray}\begin{eqnarray}
  \xdd
  &=&
  \sum_{i,j} {\mathbf C}^{-1}_{ij} P^\mathrm{d}_i P^\mathrm{d}_j
  \label{e:xdd}
  \\
  \xdc_L
  &=&
  \sum_{i,j} {\mathbf C}^{-1}_{ij} P^\mathrm{d}_i I^{(c)}_{Lj}
   \\
  \xdnu
  &=&
  \sum_{i,j} {\mathbf C}^{-1}_{ij} P^\mathrm{d}_i I^{(\nu)}_{j}
  \\
  \xdN
  &=&
  \sum_{i,j} {\mathbf C}^{-1}_{ij} P^\mathrm{d}_i I^{(N)}_{j}
\end{eqnarray}\begin{eqnarray}
  \xcc_{LM}
  &=&
  \sum_{i,j} {\mathbf C}^{-1}_{ij} I^{(c)}_{Li} I^{(c)}_{Mj}
  \\
  \xcnu_{L}
  &=&
  \sum_{i,j} {\mathbf C}^{-1}_{ij} I^{(c)}_{Li} I^{(\nu)}_{j}
  \\
  \xcN_{L}
  &=&
  \sum_{i,j} {\mathbf C}^{-1}_{ij} I^{(c)}_{Li} I^{(N)}_{j}
\end{eqnarray}\begin{eqnarray}
  \xnunu
  &=&
  \sum_{i,j} {\mathbf C}^{-1}_{ij} I^{(\nu)}_{i} I^{(\nu)}_{j}
  \\
  \xnuN
  &=&
  \sum_{i,j} {\mathbf C}^{-1}_{ij} I^{(\nu)}_{i} I^{(N)}_{j}
  \\
  \xNN
  &=&
  \sum_{i,j} {\mathbf C}^{-1}_{ij} I^{(N)}_{i} I^{(N)}_{j}
  \label{e:xNN}
\end{eqnarray}
where summation over repeated indices $L$ and $M$ is assumed in $\chi^2$.  The utility of this expression is that the $x$ coefficients in Eqs.~(\ref{e:xdd}-\ref{e:xNN}) are independent of $\vec b$.  Thus for a given model $\vec c$, the logarithm of $\mathcal L$ is a polynomial in the bias parameters.  Once these coefficients have been computed, bias marginalization can be carried out exactly for $N$ and $\btnl$, in which $\chi^2$ is quadratic.  We marginalize numerically over the remaining biases by minimizing $\chi^2$ with respect to them and then integrating numerically over intervals of width $\Delta \bd = \Delta \bdd = \Delta \bss = 2$ centered at this $\chi^2$-minimizing point.  Except where otherwise noted, we also marginalize over $\sigma_v$ as a nuisance parameter at each point in parameter space. 

\subsection*{Acknowledgments}
We are grateful to F.~Beutler, J.~Blazek, D.~Chung, X.~Fang, S.~Habib, J.~Hamann, K.~Heitmann, C.~Hirata, J.~Kwan, and Y.~Wong for insightful discussions and essential guidance.
This work was supported in part by the U.S.~Department of Energy through grant DE-FG02-95ER40896.
This research was performed using the compute resources and assistance of the UW-Madison Center For High Throughput Computing (CHTC) in the Department of Computer Sciences. The CHTC is supported by UW-Madison, the Advanced Computing Initiative, the Wisconsin Alumni Research Foundation, the Wisconsin Institutes for Discovery, and the National Science Foundation, and is an active member of the Open Science Grid, which is supported by the National Science Foundation and the U.S. Department of Energy's Office of Science.

This is an author-created, un-copyedited version of an article accepted for publication in the Journal of Cosmology and Astroparticle Physics. IOP Publishing Ltd is not responsible for any errors or omissions in this version of the manuscript or any version derived from it. The Version of Record is available online at {\tt doi.org/10.1088/1475-7516/2019/05/041}~.

\bibliographystyle{unsrt}
\bibliography{rsd_joint_analysis}

\end{document}